\documentclass[twocolumn]{aastex62}
\usepackage{amsmath,amstext,comment}

\usepackage[figure,figure*]{hypcap}
\usepackage{subfigure,graphicx} 

\usepackage{newtxmath} 
\usepackage{lmodern}
\usepackage[T1]{fontenc}
\usepackage{appendix}

\newcommand{\prot}{\ensuremath{P_{\mbox{\scriptsize rot}}}}
\newcommand{\Prot}{\ensuremath{P_{\mbox{\scriptsize rot}}}}
\newcommand{\Msun}{\ifmmode {M_{\odot}}\else${M_{\odot}}$\fi}
\newcommand{\Rsun}{\ifmmode {R_{\odot}}\else${R_{\odot}}$\fi}
\newcommand{\Lsun}{\ifmmode {L_{\odot}}\else${L_{\odot}}$\fi}
\newcommand{\lapprox }{{\lower0.8ex\hbox{$\buildrel <\over\sim$}}}
\newcommand{\gapprox }{{\lower0.8ex\hbox{$\buildrel >\over\sim$}}}
\def\amin{\ifmmode^{\prime}\else$^{\prime}$\fi}
\def\asec{\ifmmode^{\prime\prime}\else$^{\prime\prime}$\fi}

\def\Pmem{$P_{\mathrm{mem}}$}
\newcommand{\degree}{\ifmmode {^\circ}\else$^\circ$\fi}

\newcommand{\Ro}{\ifmmode {R_o}\else$R_o\ $\fi}
\newcommand{\lha}{\ifmmode {L_{H\alpha}/L_{bol}}\else$L_{H\alpha}/L_{bol}$\ \fi}

\newcommand{\Kepler}{{\it Kepler}}

\newcommand{\Gaia}{{\it Gaia}}
\newcommand{\be}{\begin{equation}}
\newcommand{\ee}{\end{equation}}

\newcommand{\msun}{M$_\odot$}

\newcommand{\mas}{\ensuremath{\mbox{mas yr}^{-1}}}

\newcommand{\kms}{\ensuremath{\rm km\,s^{-1}}}

\usepackage{xcolor}

\definecolor{my_color}{HTML}{3a18b1}

\definecolor{new_color}{HTML}{CF0000}

\newcommand{\ith}{\ensuremath{^{\rm th}}}
\newcommand{\ktwo}{\textit{K2}}

\submitjournal{\textit{The Astrophysical Journal}}

\accepted{June 15, 2021}

\shorttitle{1013 \ktwo\ Rotation Periods for Praesepe}
\shortauthors{Rampalli et al.}

\begin{document}

\title{{\sc Three K2 Campaigns Yield Rotation Periods for 1013 Stars in  Praesepe}}

\correspondingauthor{Rayna Rampalli}
\email{raynarampalli@gmail.com}

\newcommand{\columbia}{Department of Astronomy, Columbia University, 550 West 120th St, New York, NY 10027, USA}

\newcommand{\lafayette}{Department of Physics, Lafayette College, 730 High St, Easton, PA 18042, USA}

\newcommand{\amnh}{Department of Astrophysics, American Museum of Natural History, 200 Central Park West, New York, NY 10024, USA}

\newcommand{\cfa}{Center for Astrophysics $\vert$ Harvard $\&$ Smithsonian, 60 Garden St, Cambridge, MA 02138, USA}

\newcommand{\unc}{Department of Physics and Astronomy, University of North Carolina, Chapel Hill, NC 27599, USA}

\newcommand{\ut}{Department of Astronomy, The University of Texas at Austin, Austin, TX 78712, USA}

\newcommand{\wwu}{Department of Physics \& Astronomy, Western Washington University, Bellingham, WA 98225, USA}

\newcommand{\cc}{Department of Physics, Colorado College, 14 East Cache La Poudre St, Colorado Springs, CO 80903, USA}

\author[0000-0001-7337-5936]{Rayna Rampalli}
\affiliation{Department of Physics and Astronomy, Dartmouth College, Hanover, NH 03755, USA}
\affiliation{\columbia}

\author[0000-0001-7077-3664]{Marcel A.~Ag{\"u}eros}
\affiliation{\columbia}

\author[0000-0002-2792-134X]{Jason L.~Curtis}
\affiliation{\columbia} %
\affiliation{\amnh}

\author[0000-0001-7371-2832]{Stephanie T.\ Douglas}
\affiliation{\lafayette}

\author[0000-0002-8047-1982]{Alejandro N\'{u}\~{n}ez}
\affiliation{\columbia}

\author[0000-0002-1617-8917]{Phillip A.~Cargile}
\affiliation{\cfa}

\author[0000-0001-6914-7797]{Kevin R.~Covey}
\affiliation{\wwu}

\author[0000-0002-8443-0723]{Natalie M.~Gosnell}
\affiliation{\cc} 

\author[0000-0001-9811-568X]{Adam L.~Kraus}
\affiliation{\ut}

\author[0000-0001-9380-6457]{Nicholas M.~Law}
\affiliation{\unc}

\author[0000-0003-3654-1602]{Andrew W.~Mann}
\affiliation{\unc}



\begin{abstract}

We use three campaigns of \ktwo\ observations to complete the census of rotation in low-mass members of the benchmark, $\approx$670-Myr-old open cluster Praesepe. We measure new rotation periods (\prot) for 220  $\lapprox$1.3~\Msun\ Praesepe members and recover periods for 97\% (793/812) of the stars with a \prot\ in the literature. Of the 19 stars for which we do not recover a \prot, 17 were not observed by \ktwo.
As \ktwo's three Praesepe campaigns took place over the course of three years, we test the stability of our measured \prot\ for stars observed in more than one campaign. 
We measure \prot\ consistent to within 10\%
for $>$95\% of the 331 likely single stars with $\geq$2 high-quality observations; the median difference in \prot\ is 0.3\%, with a standard deviation 
of 2\%. Nearly all of the exceptions are stars with discrepant \prot\ measurements in Campaign 18, \ktwo's last, which was significantly shorter than the earlier two ($\approx$50~d rather than $\approx$75~d). This suggests that, despite the evident morphological evolution we observe in the light curves of 38\% of the stars, $\prot$ measurements for low-mass stars in Praesepe are stable on timescales of several years. A $\prot$ can therefore be taken to be representative even if measured only once.
\end{abstract}

\keywords{open clusters: individual (Praesepe) --- 
    stars:~evolution ---
    stars:~rotation ---
    stars:~late-type
    }

\setcitestyle{notesep={ }}

\section{Introduction}
\cite{skumanich72} famously found that Sun-like stars rotate more slowly as they age, and that this relationship between age and rotation can be described by a simple power law, with a star's rotation period (\prot) proportional to the square root of its age. 
In the 50 years since, many groups have used observations of populations of stars to calibrate empirically this age-rotation relation, motivated in part by the idea that such a relation could be inverted to obtain reliable ages for field stars. \cite{barnes2003} coined the term gyrochronology to describe this approach to finding ages for low-mass stars ($\lapprox$1.3~\Msun). 

\begin{figure*}[!th]
    \centering
    \includegraphics[scale=0.42]{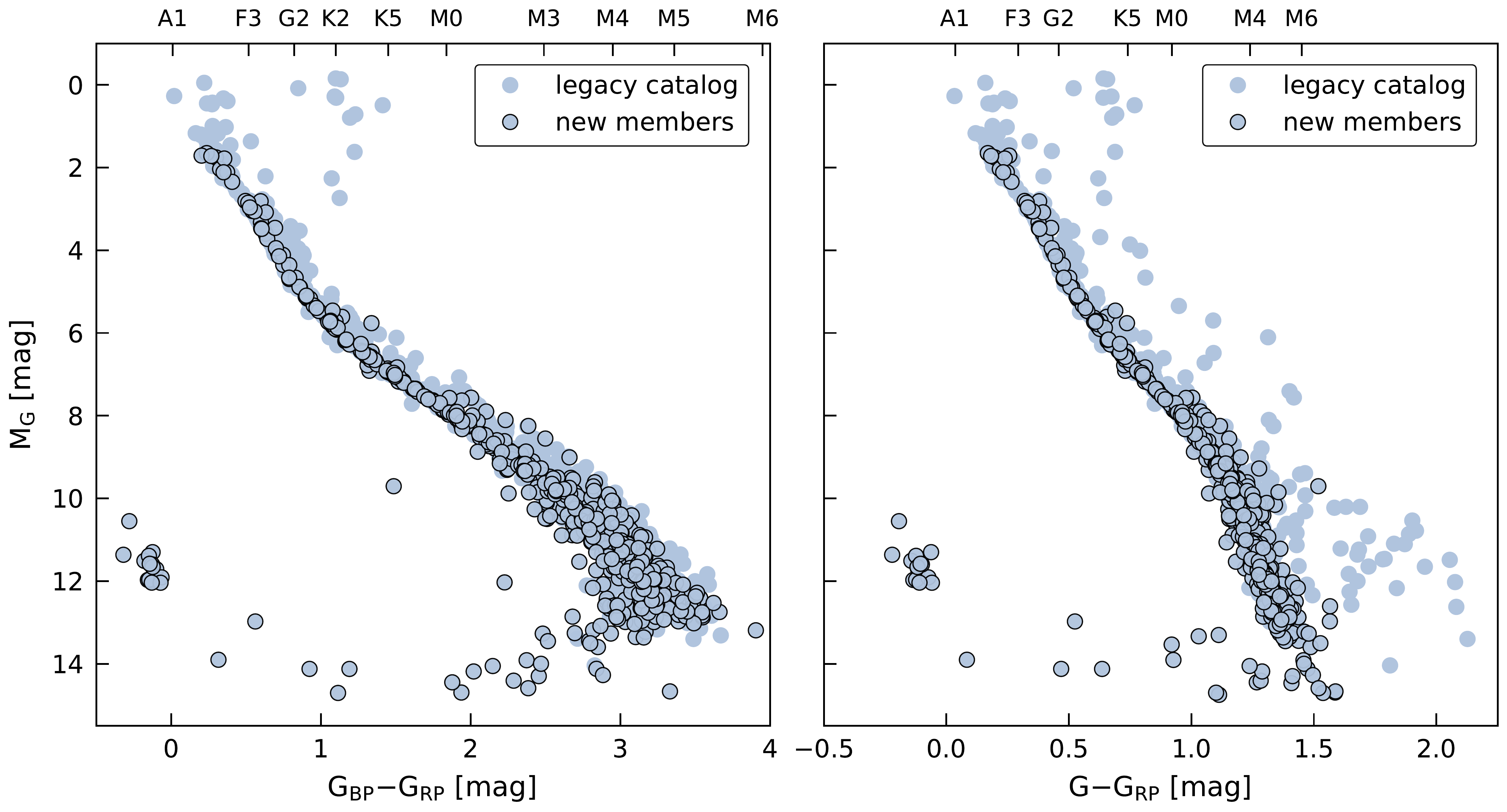}
    \caption{Praesepe CMDs based on membership catalogs assembled pre- and post-\Gaia\ DR2. The photometry has been updated to \Gaia\ EDR3 values. In both panels, we show the stars with \Pmem~> 70\% from our legacy catalog that have \Gaia\ data and add
    new members (black circles) identified with \Gaia\ DR2 data \citep{2018GaiaCluster,gaiacollab2018,Roser19,Lodieu2019}.
    \textit{Left}: G\textsubscript{BP} $-$ G\textsubscript{RP} vs. M\textsubscript{G}.
    \textit{Right}: G $-$ G\textsubscript{RP} vs. M\textsubscript{G}. 
    The stars scattered at the bottom of both panels suggest poorly measured G\textsubscript{RP} and/or  G\textsubscript{BP}. These stars do not have measured \prot\ and do not affect our analysis.}
    \label{fig:newcmd}
\end{figure*}

The last decade has proved particularly fruitful for studies of the rotational evolution of these stars, with \textit{Kepler} \citep{borucki2010}, \textit{K2} \citep{howellk2}, and most recently the \textit{Transiting Exoplanet Survey Satellite} \citep[\textit{TESS;}][]{ricker} all providing high-cadence, high-precision photometric data with which to measure \prot\ for large numbers 
of stars. This has led to extensive surveys of rotation in benchmark open clusters such as the Pleiades \citep[e.g.,][]{rebull2016-1,rebull2016} and the Hyades \citep[e.g.,][]{douglas2016,douglas2019},
and more recently, in stellar streams and moving groups \citep[e.g.,][]{curtispier,mannursamajor}.

We revisit another  benchmark open cluster, 
Praesepe \citep[$\approx$670 Myr;][]{douglas2019}. Recent \textit{Gaia} data releases have significantly improved our knowledge of the membership of this and other open clusters \citep[cf.][]{2018GaiaCluster}. For example,  \citet{Roser19} and \citet{meingast2019} used these data to detect tidal tails in several clusters. These authors have identified new high-confidence members that can be matched to existing photometric data, thereby potentially increasing the number of rotators even for clusters like Praesepe that have already been extensively surveyed for \prot.   

Furthermore, $\ktwo$ observed Praesepe during three separate campaigns (C5, C16, and C18, $\ktwo$'s last campaign). \citet[; hereafter D17]{douglas2017} published \prot\ measured from C5 data, but \prot\ have not been published using data from C16 and C18. And because many stars were observed in at least two of these campaigns, combining the data from the three campaigns provides a rare opportunity to examine the stability of light curves for a large number of low-mass stars over, in a few hundred cases, nearly three years.

We update the \prot\ catalog for Praesepe given our new membership catalog and the addition of two campaigns-worth of $\ktwo$ data. We also show that over the three years that separate C5 and C18, the vast majority of rotators observed by $\ktwo$ more than once have \prot\ measurements that change by no more than a few percent. This suggests that, despite the significant morphological evolution we observe in the light curves of 38\% of the stars, $\prot$ measurements for low-mass stars in Praesepe are stable on timescales of several years, and can be taken to be representative even if measured only once.

In Section~\ref{section:membership}, we describe our new Praesepe membership catalog, which combines our pre-$\Gaia$ catalog with  new $\Gaia$ astrometric measurements, and contains 1708 stars with a membership probability \Pmem~$> 70\%$. 
In Section~\ref{section:methods}, we examine the existing \prot\ data for Praesepe in light of our new membership catalog. We then measure and validate new \prot\ using $\ktwo$ photometry. In Section~\ref{section:discussion}, we discuss our results. We conclude in Section~\ref{section:conclusion}.

\section{Revisiting the Praesepe Membership Catalog in the era of \Gaia}\label{section:membership}

\subsection{A New \Gaia~DR2-based Membership Catalog}
Our original Praesepe catalog, the legacy catalog, described in \cite{douglas2014} and D17, is based on the \cite{kraus2007} membership catalog, which calculated \Pmem\ using photometry and proper motions from a number of surveys. An additional 39 previously identified members, too bright to be identified by those authors, were added as {\it bona fide} members. The legacy catalog includes 1092 stars with \Pmem~> 70\%. 

After the release of \Gaia\ DR2, we revised this catalog using the membership catalogs of  
\cite{2018GaiaCluster}, \cite{gaiacollab2018}, \cite{Roser19}, and \cite{Lodieu2019}. The first two of these new catalogs focused on members in the cluster core, whereas the latter two catalogs also found members outside of Praesepe's tidal radius (10.77 pc;  \citealt{Roser19}), in the tidal tails and in a halo around the cluster.

With the \Gaia\ DR2 catalogs, we recover 95\% (1059/1092) of the  high confidence members that were in our legacy catalog. The 33 stars that we do not recover are more than 3.5x the tidal radius away from the center of the cluster and are not included in any of the \Gaia-based catalogs that identify the tidal-tails or halo.\footnote{While this suggests these stars are too far from Praesepe's core to be cluster members, we find \prot\ for 13 of these and include them in Tables \ref{t:ls} and \ref{t:prot}.} N\'{u}\~{n}ez et al. (in prep) will investigate these stars further.

We add 616 previously unidentified members from the \Gaia\ catalogs. Since many of these catalogs do not include membership probabilities, in the analysis that follows, we treat each of these candidates as a \Pmem $> 70\%$ member. 

With these additions, our updated catalog includes
1708 stars that have \Gaia\ data and with \Pmem~> 70\% , the threshold we use for our \prot\ analysis for consistency with our previous studies (\citealt{douglas2014}, D17). Of these stars, we find approximately 730 stars to be outside of the cluster's tidal radius; 318 of these stars were identified by \cite{Roser19} to be in Praesepe's tidal tails, which are defined to be found beyond 2.5x the tidal radius. 

\subsection{The Impact of \Gaia~EDR3 on Our New Catalog}
The work described above was completed prior to the \Gaia\ Early Data Release 3 \citep[EDR3;][]{edr3}. Gaia EDR3 identified $\approx$10$^8$ more sources than DR2, and improved the precision of parallaxes, proper motions, and photometry by $\approx$30\% for stars already included in DR2. 

We update the astrometry and photometry to the EDR3 values for our entire catalog, regardless of membership probability. We also make additional corrections to the magnitudes as recommended in Appendix~C of \cite{riello20} and use EDR3 distances from \cite{edr3dists}. Three stars not in DR2 have EDR3 observations. In addition, $\approx$60\% of the stars that did not have DR2 parallaxes have EDR3 parallaxes, and $\approx$30\% of the stars that did not have DR2 proper motions now have proper motions. 

Forty-three stars in our legacy catalog, while resolved by \Gaia\, lack proper motions and parallax measurements, including 23 stars in our rotator sample. 
 
For these 23, we assign a parallax $\pi = 5.95$$\pm$$0.40$~mas, the error-weighted mean parallax of all the cluster members, $D = 183.2$$\pm$$13.45$pc, the error-weighted mean distance of all the cluster members, and use the proper motions reported for these stars in \cite{kraus2007}.

Figure~\ref{fig:newcmd} shows two color-magnitude diagrams (CMDs) illustrating the difference between the legacy catalog and our expanded, improved catalog. The complete membership catalog will be available in N\'{u}\~{n}ez et al. (in prep).  

The most significant impact of this update to EDR3 values on our work is on our identification of candidate binaries in the cluster, as discussed below.

\subsection{Binary Identification}\label{section:binaries}
The rotation evolution of members of binary systems can differ significantly from that of their single-star counterparts \citep[e.g.,][]{meibom2005, zahn2008}. For example, stars in tight binary systems can exert tidal forces on each other,  causing the stars to spin up or down and deviate from the spin-down evolution expected for a single star. 
Furthermore, depending on the tightness of the binary and the resolution of the detector, a companion may contaminate the rotational signal we are attempting to measure from a star.\footnote{This is also true of background or foreground stars, of course, but these are on average far older than stars in Praesepe and therefore much less likely to have light curves with strong rotational modulations.} Either of these situations can lead to stars whose positions in the color-period plane are incorrect, which in turn undermines our ability to compare accurately the \prot\ distribution of clusters of different ages and to explore evolutionary trends. 

We use the high-precision \Gaia\ EDR3 astrometry and \ktwo\ photometry and follow \cite{douglas2019} in formulating the following tests to identify binaries among the cluster members with $P_{\rm mem} > 70\%$: 

\begin{enumerate}
    \item 
    The orbital motions of stars in tight binary systems typically affect the individual stars' proper motions ($\mu$) such that these differ significantly from the cluster's mean proper motion, $\bar \mu$. Based on our membership catalog, we calculate  $\bar \mu_\alpha = -35.95$ and $\bar \mu_\delta = -12.90$ mas yr\textsuperscript{-1} for Praesepe, and flag any star with a radial proper motion $\geq$2.5~mas~yr\textsuperscript{-1} different from the corresponding $\bar \mu$.\footnote{We choose not to apply this test to stars in the cluster's tidal tails, as these naturally have large proper motion deviations from the cluster mean. These stars are only flagged as binaries if they meet one of the other conditions described here.} 
    \\ 
    We also examine the renormalized unit weight error (RUWE) measurement for each star. This is a goodness-of-fit measure of the single-star model fit to the source's astrometry. If a star has a RUWE~$> 1.2$, there is a strong likelihood the star has a companion, and the system is usually a wide binary (e.g., \citealt{ruwe2,ruwe1}).  
    \item 
    Equal-mass binaries appear over-luminous and sit above the single-star main sequence in a CMD. As was done in \cite{douglas2019}, we fit the single star sequence in D17 (updated with EDR3 values) with a sixth-order polynomial for both the  G\textsubscript{BP} $-$ G\textsubscript{RP} vs. M\textsubscript{G} and the G $-$ G\textsubscript{RP} vs. M\textsubscript{G} CMDs. If stars in our new catalog are $\geq$0.375~mag brighter than the polynomial fit's G for stars of the same color, we flag them as candidate binaries. Since the position of binaries relative to single stars becomes  harder to disentangle at redder colors, we only flag stars if they are bluer than G\textsubscript{BP} $-$ G\textsubscript{RP} = 2.65 or G$-$G\textsubscript{RP}= 1.15 mag. In the G$-$G\textsubscript{RP} vs. M\textsubscript{G} CMD, we also flag obvious outliers if they are $\geq$5$\times$ the 0.375 mag difference away from the single star sequence. No such outliers are seen in the G\textsubscript{BP} $-$ G\textsubscript{RP} vs. M\textsubscript{G} CMD. 
    \item
    For stars that have a radial velocity (RV) in \Gaia\ DR2 or in other RV surveys, we compare this RV to the mean for the cluster, 35~\kms\ \citep{clusterparams}. We flag any star with an RV~>~2~\kms\ away from the mean cluster RV or with an RV error $>$2~\kms.\footnote{There are 11 rotators that are flagged as candidate binaries solely due to their \Gaia\ RV error. While the errors are larger than our adopted threshold, they are all under 4~\kms, so it is not clear whether they are short-period binaries. The classifications for those fainter than G$>12$ mag should also be treated with caution, as the Gaia RV error increases with G due to signal-to-noise issues and not actual RV variations.} 
\end{enumerate}

\begin{figure}[!t]
    \centering
    \includegraphics[scale=0.38]{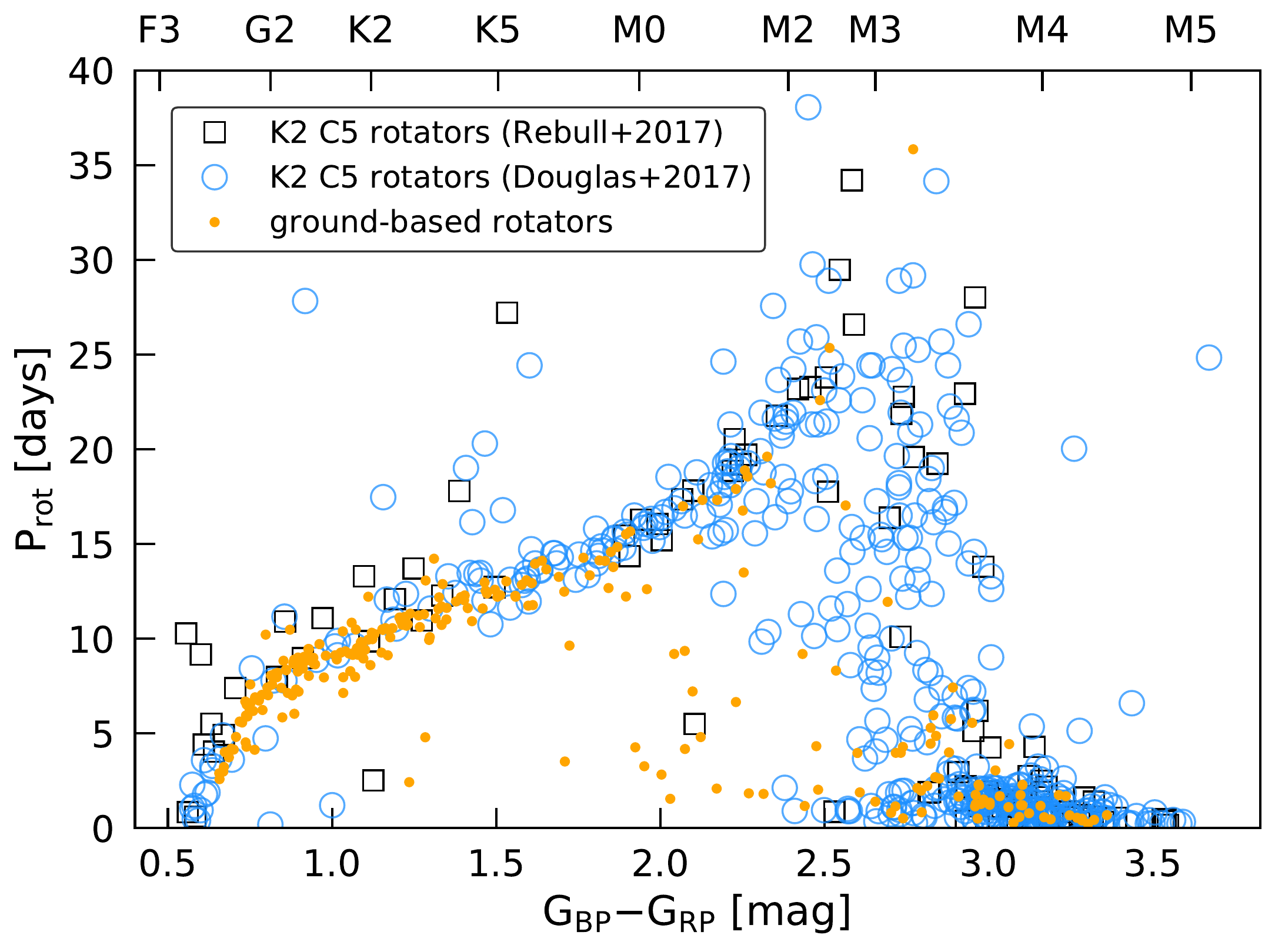}
    \caption{\Gaia\ color-period diagram for rotators identified prior to this work. Yellow dots are from the ground-based observations of  \cite{scholz2007}, \cite{scholz2011}, \cite{delorme2011}, \cite{agueros11}, and
    \cite{kovacs2014}. The blue circles indicate \Prot\ measured by D17 using \ktwo\ C5 data. The black squares indicate \Prot\ measured by \cite{rebull17} in C5 and not measured by D17.}
    \label{fig:oldrots}
\end{figure}

Failing any one of these  tests is sufficient to be labeled a candidate binary. We find that  $\approx$50\% of the stars in our catalog are either candidate or confirmed binaries. The final stage of our binary-identification process is specific to the stars for which we measure a \prot\ and is described in Section~\ref{sec:final}.

\section{Completing the \prot\ Census of Praesepe}
\label{section:methods}

\subsection{Rotators From the Literature}
Prior to the start of the \ktwo~mission, there were 269 published low-mass rotators in Praesepe. These measurements came from a variety of ground-based observations, described notably in \cite{scholz2007}, \cite{scholz2011}, \cite{delorme2011}, \cite{agueros11}, and \cite{kovacs2014}. Using our membership catalog, we find that 262 of these 269 rotators have a \Pmem~> 70\%. 

In addition to recovering many of these periods, D17 added $475$ new \prot\ using C5 observations. Four-hundred and fifty stars have a \Pmem~> 70\%. 
\citet[; hereafter R17]{rebull17} also used C5 to measure \prot\ for Praesepe members, and reported \prot\ for 809 stars. Of these stars, 445 have \prot\ that were also measured by D17, and $>$95\% have periods that agree within 10\%.\footnote{Since R17 and D17 used the same \ktwo\ campaign for finding \prot, discrepancies in the reported \prot\ represent systematic differences in the measurement methods.} An additional 205 had previous ground-based observations; $>$93\% of these have \prot\ in agreement. Of the 159 remaining stars, we find that 100  have a \Pmem~$>$~70\% and meet our G\textsubscript{BP} $-$ G\textsubscript{RP} > 0.55 mag cutoff, which we use to identify the $\lapprox$1.3~\Msun\ stars in our catalog. 

Before measuring any \prot\ in this work, we therefore assemble a sample of 812 rotators with \Pmem~> 70\%. The \Gaia\ color-period distribution of these literature measurements is shown in Figure~\ref{fig:oldrots}.

\subsection{Post-C5 \ktwo\ Observations of Praesepe}
\begin{figure*}[!th]
    \centering
    \includegraphics[scale=0.28]{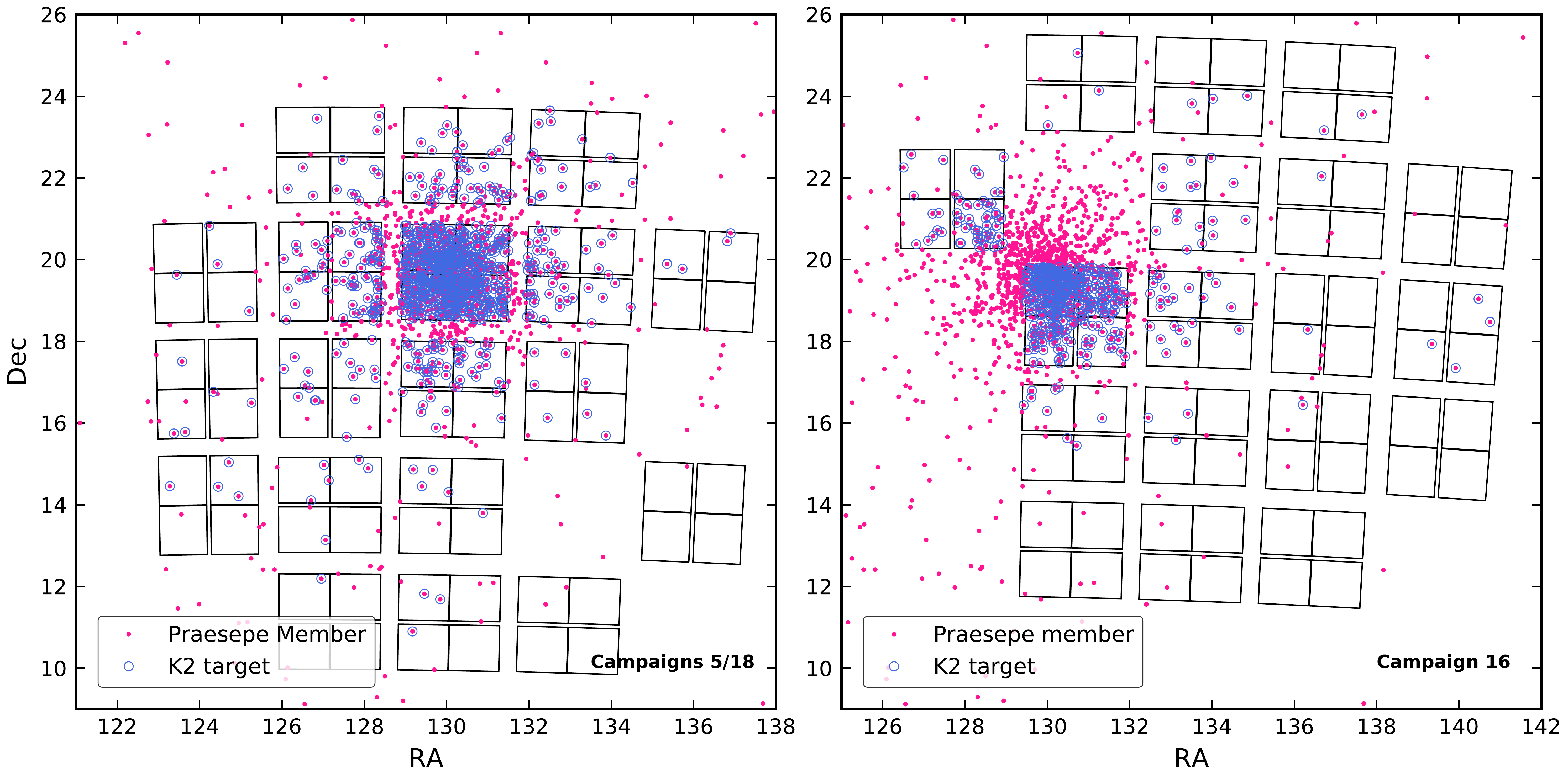}
    \caption{\ktwo\ observations of Praesepe. Praesepe members with \Pmem~> 70\% are shown as pink dots; those with \ktwo\ data are highlighted with blue circles. Three hundred and twenty-seven members were observed in all three campaigns. \textit{Left:} The aimpoints for C5 and C18 were almost identical. Eight hundred and sixty-eight stars have light curves from both campaigns. \textit{Right:} The field-of-view for C16, which was offset from the cluster core. Three hundred and thirty-two Praesepe members had C5 and C16 light curves; 340 had C16 and C18 light curves.}
    \label{fig:k2fp}
\end{figure*}

Praesepe was a \ktwo\ target during three campaigns conducted over three years  (see~Table~\ref{tbl:k2obs}). For the first, C5, the aimpoint was set to maximize coverage of the cluster, with the cluster core being placed on four of the detector CCDs. For the first of its returns, C16, the \ktwo~aimpoint was offset from that of C5 by several degrees. By contrast, C18 was designed to be an almost exact copy of C5 (see Figure~\ref{fig:k2fp}).

We successfully proposed Praesepe targets for observation during C16 and C18 (Proposals  K2GO52-0060 and  K2GO6-0040). Our target lists were constructed before \Gaia\ DR2, and were therefore based on the legacy catalog. For completeness, we also included cluster stars in that catalog with \Pmem\ as low as 10\%, and for C18, we added 428 M-dwarf members identified from UKIDSS data \citep{boudrealt2012}. After comparing the list of proposed stars from all three campaigns to the stars in our new catalog, we found 951 stars in common and downloaded light curves for 219 additional stars identified in the new \Gaia\ catalogs. In total, we download 1170 \ktwo\ light curves out of the 1708 (68.5\%) high-confidence members in our membership catalog. 

\begin{deluxetable}{cccc}
\centering 
\tabletypesize{\footnotesize} 
\tablecolumns{4}
\tablewidth{0pt}
\tablecaption{\ktwo~observations including Praesepe}
\label{tbl:k2obs}
\tablehead{
\colhead{Campaign} & 
\colhead{Start Date} & 
\colhead{Length (days)} & 
\colhead{Field Center}
}
\startdata
5 &	2015 Apr 27 & 74 & 08:40:38 +16:49:47\\ 
16 & 2017 Dec 07 & 80 & 08:54:50 +18:31:31 \\
18 & 2018 May 12 & 51 & 08:40:39 +16:49:40 
\enddata
\tablecomments{Because the spacecraft was running out of fuel $\approx$50 days into C18, the collection of science data was terminated earlier than scheduled. The spacecraft was put in a hibernation-like state until the data were downloaded in early August 2018. C18 was the last \ktwo~campaign. }
\end{deluxetable}

\begin{deluxetable}{ccccc}
\centering 
\tabletypesize{\footnotesize} 
\tablewidth{0pt}
\tablecaption{Cluster members observed by \ktwo}
\label{tbl:overlap}
\tablehead{
\colhead{Campaign} & 
\colhead{Members} & 
\colhead{Also in C5} & 
\colhead{Also in C16} &
\colhead{Also in C18} 
}
\startdata
5 &	911 & \nodata & 332 & 868 \\ 
16 & 512 & 332 & \nodata & 340 \\
18 & 894 & 868 & 340 & \nodata
\enddata
\tablecomments{A total of 327 stars were observed in all three campaigns.}
\end{deluxetable}

Table~\ref{tbl:overlap} summarizes the number of cluster members with \Pmem~> 70\% and G\textsubscript{BP} $-$ G\textsubscript{RP} > 0.55 mag ($\lapprox$1.3~\Msun) observed in each campaign. Out of the 1170 stars in our membership catalog with \Pmem~> 70\% and \ktwo\ data, we find 1104 meet this threshold and analyzed light curves these unique targets observed over the three campaigns. The number of stars with repeat observations is significant, particularly between C5 and C18 (868 stars). Three hundred and twenty-seven stars were observed in all three campaigns.

\subsection{\textit{K2} Light Curves}
When the \textit{Kepler} mission was recommissioned as \textit{K2},
it was in an unstable equilibrium against solar pressure. The spacecraft drifted and had its thrusters fire every 6 hr to return it to its original position. As a result, stars moved in arcs across the focal plane, and the resulting light curves have a characteristic saw-tooth pattern \citep{vancleve2016}. To account for this effect, we initially applied the \textit{K2} Systematics Correction  \citep[K2SC,][]{aigrain2016} software to the mission's light curves. As discussed in D17, this approach seemed the best for removing systematics and long-term trends that can mask the periodic signals of interest to us. 

In the interim, however, several campaigns of the mission's pre-search data conditioning (PDC) light curves \citep{vancleve2016} were reprocessed. These were released along with the pre-search data conditioning simple aperture photometry (PDCSAP) light curves, which were corrected for the telescope systematics \citep{smith12,stumpe12}. 

We tested our period-detection algorithms, described below, on both the K2SC and the PDCSAP light curves, finding virtually no difference in our results. We chose to use the PDCSAP light curves for all of the stars to maintain uniformity in our analysis. The only additional processing we did was to perform a sigma clip, removing points $\geq$5$\sigma$ from the light curve mean.

\subsection{Measuring Rotation Periods}
Our approach to measuring \prot\ follows that described in D17. Using the \cite{press1989} FFT-based Lomb--Scargle (LS) algorithm\footnote{The LS approach to searching for periods is computationally straightforward to implement as it can be applied directly to  data without interpolation. While this is not an issue for \ktwo\ data, it is for {\it TESS}, whose light curves feature large gaps. Thus, LS is likely to remain the preferred technique for quickly determining \prot\ from {\it TESS} light curves. For this reason, we focus our analysis on LS-derived periods to ensure that our results will be maximally relevant to upcoming {\it TESS} studies.}, we compute the periodogram power for periods ranging from 0.1-40 d for C5 and C16, and 0.1-30 d for C18. The upper limit corresponds to a bit more than half the length of the \textit{K2} campaigns.

To assess the robustness of the \prot\ we measure, we follow \cite{ivezic2013} and use a normalized power $P_{LS}$ for each periodogram. The closer $P_{LS}$ is to 1, the more likely the signal is sinusoidal, as opposed to noise. Rather than imposing a global minimum value for $P_{LS}$, we compute a minimum significance threshold using bootstrap re-sampling, as in \cite{douglas2016}. Holding the observation epochs fixed, the flux values are randomly redrawn and replaced to create new light curves for which new periodogram powers are calculated. This process is repeated 1000 times, and the 99.9\ith\ percentile peak power is the minimum threshold for that periodogram. A peak in the original light curve is only significant if the power is higher than this threshold and  higher than at least 100 of its neighboring points. The highest of these significant peaks is taken as the \Prot. 

\begin{figure*}
    \centering
    \includegraphics[width=\textwidth,height=\textheight,keepaspectratio]{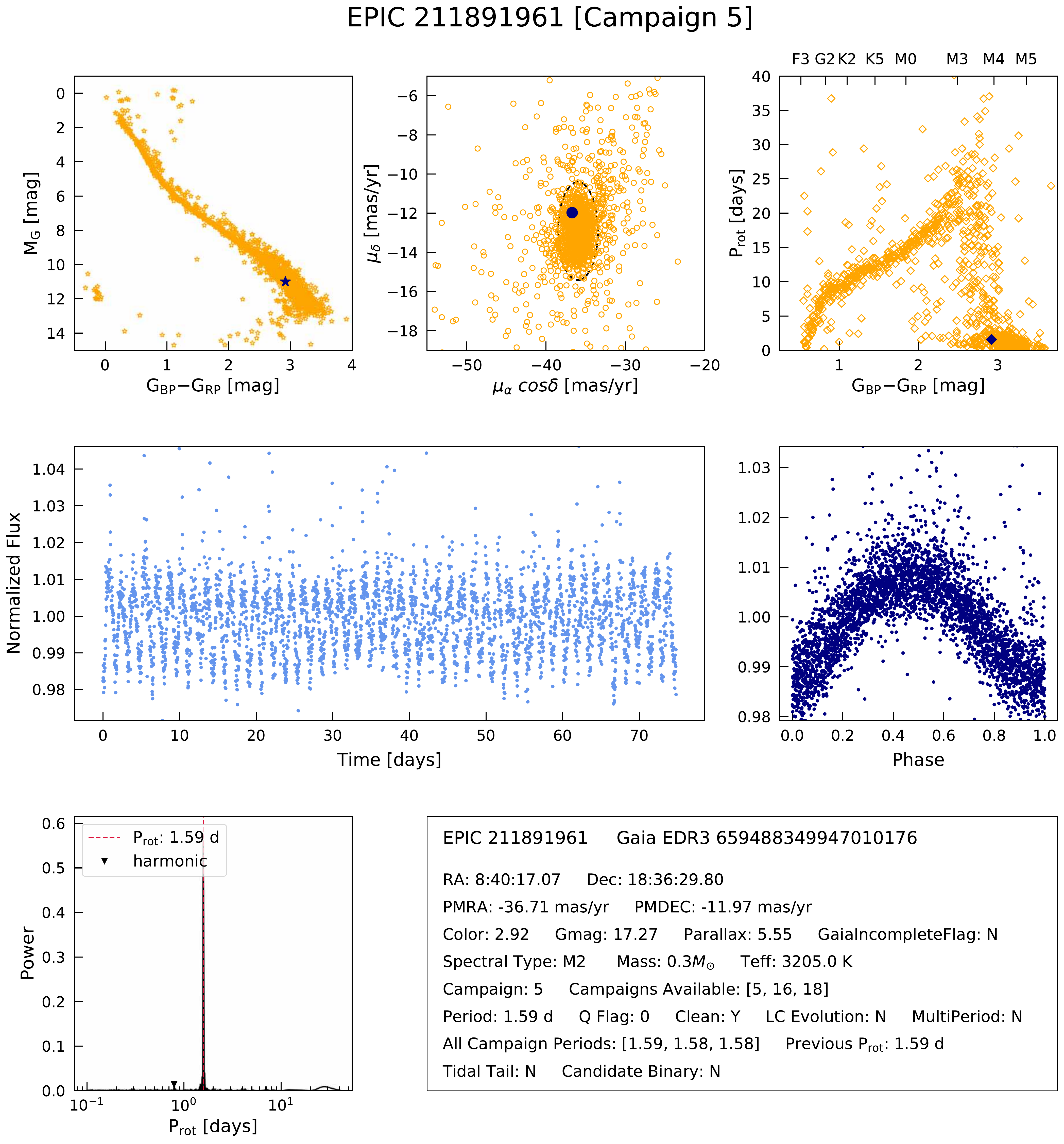}
    \caption{Example figure set for a star for which we measure a confident \prot\ (marked as clean and with Q = 0 or Q = 1), EPIC 211891961. \textit{Top left:} \Gaia\ CMD of Praesepe members (orange stars) with star denoted in navy. \textit{Top middle:} \Gaia\ proper motion distribution of Praesepe members (orange circles) with star in navy. \textit{Top right:} Color-period distribution of Praesepe rotators (orange diamonds) with star's in navy. \textit{Middle left:} Light curve for this campaign. \textit{Middle right:} Phase-folded light curve using the \prot\ measured for this campaign. \textit{Bottom left:} Lomb-Scargle periodogram. \textit{Bottom right:} Information about the star, including EPIC ID, \Gaia\ EDR3 name, right ascension, declination, proper motion, \Gaia\ color, G magnitude,  parallax, missing \Gaia\ astrometry flag, spectral type, mass, effective temperature, campaign for this light curve, campaigns observed, \prot\ measured, quality flag assigned, clean detection flag, light-curve evolution observed, multiple periods observed, all periods measured, previous \prot\ measured, tidal-tail status, and binary status. The complete figure set is available in the online journal.
}
    \label{fig:diag}
\end{figure*}

\subsection{Validating Rotation Periods}
\label{section:validation}
Following \cite{covey2016} and D17, our validation of the measured \prot\ includes two further checks, one automated and one by eye. We define a clean \prot\ detection as one where there are no other periodogram peaks with more than 60\% of the primary peak's power. Otherwise, we flag the detection as not clean to indicate that we are less confident in the measured \prot. 

Using a quality check system, we then inspect the light curves and periodograms and assign Q = 0 for obvious detections, Q = 1 for questionable detections (usually when there are obvious modulations in the light curve but an unclear \prot), Q = 2 for spurious detections, and Q = 3 for cases where the light curve is dominated by systematics or there is no significant peak in the periodogram. Thus, high-quality \prot\ measurements are those classified as clean and with Q~=~0. We illustrate the process of assigning quality flags with a set of example light curves and periodograms in  Figure~\ref{fig:qfex} in Appendix~\ref{sec:qfd}. 

We identify 64 stars for which all of the \prot\ measurements have a Q $= 1$. We choose to include this small sample of stars along with the Q $=0$ stars in our final sample to provide a more complete rotator sample.

We flag 106 cases where more than one period was detected.
We use this flag when we see multiple periods in the light curve(s) that can phase to Q = 0 quality and are non-aliases of one another. This is usually indicative of a binary system, and these stars are added to the list of candidate and known binaries discussed in Section \ref{sec:final}. 

Finding multiple periods in a light curve could also be the result of having multiple stars in the \ktwo\ aperture. To disentangle this effect from that of binarity, we check the \ktwo\ target pixel files (TPFs) for these 106 stars. Using \Gaia, we look for nearby stars with a magnitude $\leq$ the target star's magnitude + 1 mag. We find 12 stars that have multiple stars in the TPF. Upon visual inspection, however, there are no obvious cases where the \ktwo\ pipeline aperture encompasses both of the stars. To be conservative, we flag five of these 12 stars as having potential contamination from the neighboring star (see Table \ref{t:ls}). These five stars had already been flagged as candidate binaries through other tests. This suggests that it is more likely that the cases in which we detect multiple periods are indeed more likely binaries unresolved by \Gaia.

Figure~\ref{fig:diag} is an example of the figure set we produce for every Praesepe star with a confident (clean, Q=0 or Q=1) \prot\ measurement in any of the three \ktwo\ campaigns. %
The top row provides membership information: the star's location on a CMD, in proper motion space, and in the color-period plane.

The middle row shows the light curve for that campaign and the phase-folded light curve given the measured period. 
The bottom row shows the periodogram and a table with relevant information about the star.

\subsection{Assigning Final Rotation Periods}\label{section:finalprot}
Most of the stars for which we measure \Prot\ have light curves from multiple \ktwo\ campaigns. If we flagged the \prot\ from each campaign as clean and having Q $= 0$, we assign the star a \prot\ equal to the median period. Otherwise, we take the period from the campaign(s) that has (have) the cleanest detection(s) and/or best quality flag(s).

We identify 234 stars where periods measured in at least one campaign are the half-period harmonic instead of the true \Prot, likely as a result of symmetrical spot configuration and/or spot evolution on the stellar surface \citep{mcquillan2013} in at least one campaign. Occasionally, there are clean detections with Q $= 0$ that have a \prot\ that is half of the \prot\ found in other campaigns (which may have had unclean detections or worse quality flags). We doubled the aliased \Prot\ and use this corrected value to obtain the stars'  final \prot. We also flag these stars as having harmonics.

\begin{figure*}[th!]
    \centering
    \includegraphics[width=\textwidth,keepaspectratio]{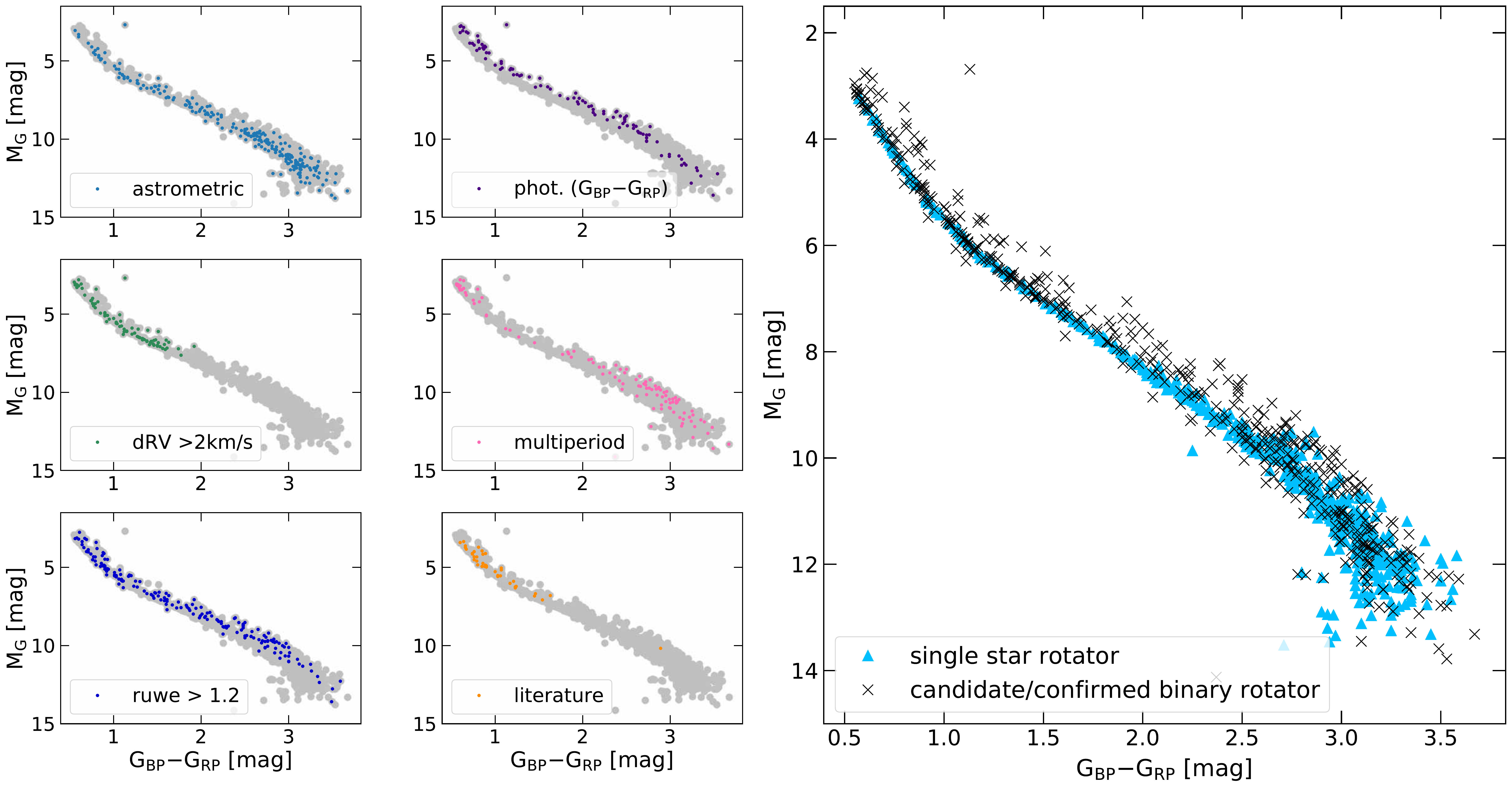}
    \caption{Binaries identified by our various tests using \Gaia\ and \ktwo\ data and from the literature. \textit{Top left:} The blue circles are stars flagged as candidate astrometric binaries because dPM $\geq 2.5$ mas yr\textsuperscript{-1}. \textit{Top middle:} Stars flagged as candidate photometric binaries because their M\textsubscript{G} $\geq 0.375$~mag than the M\textsubscript{G} for a star on the single-star sequence (this was done for both G\textsubscript{BP} $-$ G\textsubscript{RP}, the purple circles, and G$-$G\textsubscript{RP}, the red circles). \textit{Middle left:} The green circles are stars flagged because dRV or the RV error $\geq 2$~km s$^{-1}$. \textit{Middle middle:} The pink circles are stars flagged because one or more of their light curves exhibits multiple periodic signals. \textit{Bottom left:} The dark blue circles are stars flagged as candidate wide binaries because their RUWE $\geq 1.2$. \textit{Bottom middle:} The orange circles are stars reported as binaries in the literature. \textit{Right:} Resulting CMD with all known/candidate binaries shown with black crosses and likely single-star rotators with light blue triangles.}
    \label{fig:binaries}
\end{figure*}

In Table \ref{t:ls}, we provide all of the LS outputs for each star and for each campaign. We report the primary period, the primary power, the secondary period, the secondary power, the assigned quality flag, if there was a clean detection of the period, the minimum significance threshold, any harmonics, the amplitude (R\textsubscript{var}), the signal-to-noise ratio (SNR) of the \ktwo\ light curve, and any prior \prot\ measured for the star. We measure R\textsubscript{var} using the difference between the 5$^{\rm th}$ and 95$^{\rm th}$ percentiles of the (sorted) light-curve flux.

We also include a flag for light-curve evolution. We use this flag to indicate  changes in morphology within the light curve from a single campaign and/or from one campaign to another.\footnote{Since the \ktwo\ pointing and systematics are significantly less stable than \Kepler's, it is difficult to use R\textsubscript{var} as a quantitative metric for tracing spot evolution, as was done in \cite{mcquillan2014}. After unsuccessfully testing a number of quantitative metrics to trace light-curve evolution, we check each light curve by eye.} These changes could be indicative of spot evolution and/or differential rotation. The results of this flagging are discussed in Section \ref{section:spotev}.

We measure \prot\ for 220 new stars. We also recover rotation periods for $>$97\% (793/812) of the stars with measured \prot\ in the literature, resulting in 1013 rotation periods from \ktwo\ observations. Of the 19 stars we did not recover, two have a \Prot\ in R17. In our analysis, however, the \Prot\ we found did not pass our validation process. The remaining 17 stars have a ground-based measurement of their \prot\ but were not observed by \ktwo. We include these 17 in our final catalog of rotators.

We now have \prot\ measurements for a total of 1030 stars in Praesepe. Since there are 1564 stars with masses $\lapprox 1.3$~\Msun\ in our membership catalog, we have \prot\ for $\approx$63\% of the low-mass stars in Praesepe, a remarkable total given that \ktwo\ observed $\approx$69\% of all of the stars in our catalog.  We show the color-period distribution of this sample in Figure~\ref{fig:cpdnew} and provide the stars' properties in Table~\ref{t:prot}. 

\subsection{Assembling the Definitive Praesepe Rotator Catalog}\label{sec:final}

\subsubsection{Mass \& Temperature Estimations} 
The \Gaia\ G\textsubscript{BP} $-$ G\textsubscript{RP} color is a reasonable proxy for effective temperature (e.g., \citealt{Casagrande2020,curtisrup14720}). We linearly interpolate G\textsubscript{BP} $-$ G\textsubscript{RP} and temperatures from \cite{MamajekTable}\footnote{We use the table available at \url{http://www.pas.rochester.edu/~emamajek/EEM_dwarf_UBVIJHK_colors_Teff.txt}.} and use this function to infer temperatures for the rotators in our sample. 

As advocated by \citet{constrainmdwarf}, we use absolute $K$ magnitudes to estimate our rotators' masses. We calculate $M_K$ from  apparent 2MASS or UKIDSS $K$ magnitudes  \citep{2mass,boudrealt2012} and \Gaia\ distances \citep{edr3dists}. In this conversion, we use the extinction value of E(B-V)$ =0.035$ from \cite{douglas2019} multiplied by a coefficient of 0.302 from \cite{schlafly} assuming an extinction factor of $R_{v} = 3.1$ to find the extinction in the $K$ band. As with effective temperature, we linearly interpolate $M_{K}$ and mass from \cite{MamajekTable} and determine the masses for our stars from this relation. 

Using this method of mass determination, we find that 17 of the rotators have $>$1.3~\Msun, beyond the mass range in which stars experience solar-like spin-down and 
for which gyrochronology is applicable. Since these stars are all candidate or confirmed binaries, it is likely they appear overluminous in the K-band and thus have overestimated masses. We test this by increasing $M_K$ by 0.5~mag for our single-star sequence and calculating the corresponding fractional increase in inferred masses compared to the original values. This results in an average fractional increase of 20\% across the mass range of this sample, so it is unsurprising to have a handful of stars that are likely binaries with an inferred mass $>$1.3~\Msun. 

\subsubsection{More Binary Identification}
In addition to the binary determination process outlined in Section \ref{section:binaries}, we flag our rotators as candidate binaries if we find multiple periods in any of their \ktwo\ light curves, or if there is a confirmation of their binarity in the literature. We have 106 cases of stars with multiple periods in their light curves and 39 stars described as binaries in the literature. 

In total, we find 473 candidate or confirmed binaries in the sample of 1030 Praesepe rotators: 236 stars with a proper motion deviation greater than 2.5 mas/yr (astrometric binaries), 176 stars with a RUWE $\geq 1.2$ (likely wide binaries), 100 photometric binaries, and 85 stars that have discrepant RV measurements. This rate of binarity ($\approx$50\%) is comparable to that in D17. 

\begin{figure*}
   \centering
    \includegraphics[scale=0.65]{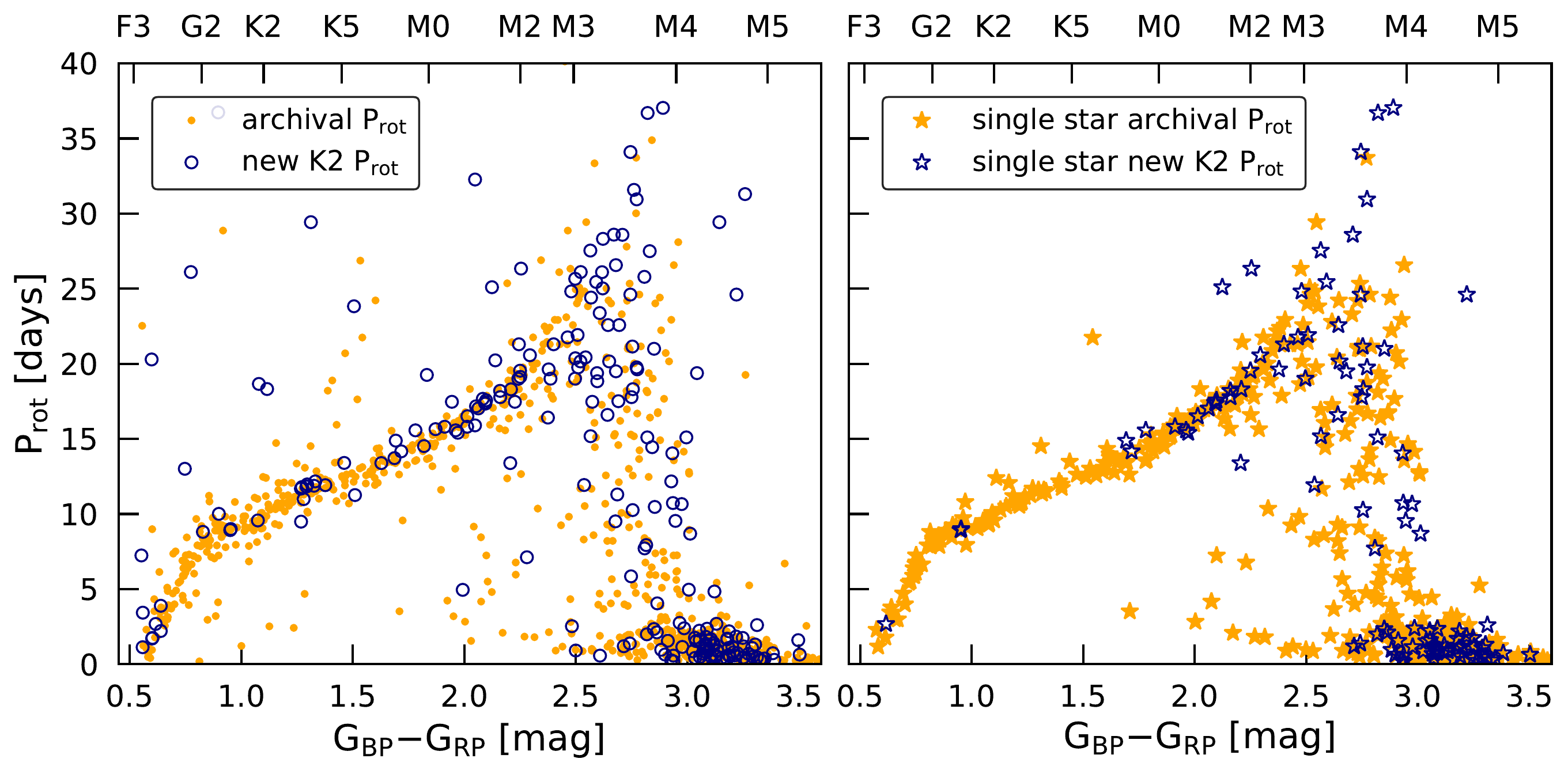}
    \caption{\Gaia\ color-period distribution for Praesepe rotators. Rotators found prior to this work are in orange. New rotators are in navy. \textit{Left:} The entire rotator sample with likely binaries included.  \textit{Right:} The likely single-star rotator sample.}
    \label{fig:cpdnew}
\end{figure*}

\begin{figure*}
    \centering
    \includegraphics[scale=0.65]{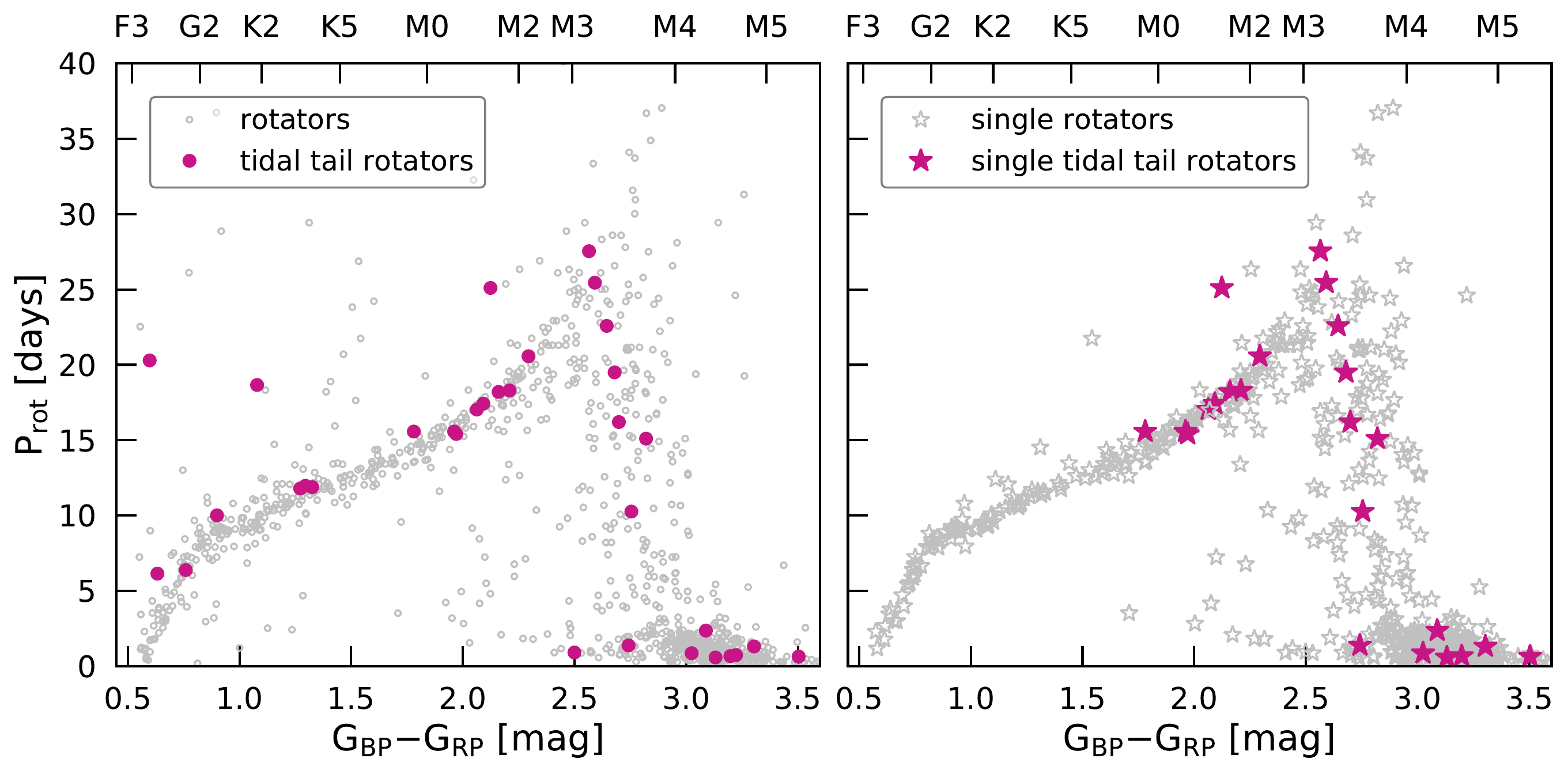}
    \caption{\Gaia\ color-period distribution for rotators in Praesepe's tidal tails. Rotators in the cluster core or outside the tidal radius, but not in the tails, are in grey. Tidal tail stars are in magenta. \textit{Left:} Entire rotator sample with likely binaries included. \textit{Right:} Likely single rotator sample.}
    \label{fig:cpdtt}
\end{figure*}

\indent CMDs highlighting the positions of these candidate and confirmed binaries as identified by our different tests are shown in Figure~\ref{fig:binaries}, and we use this information to generate color-period plots featuring only the likely single stars (right panels of Figures \ref{fig:cpdnew} and \ref{fig:cpdtt}). Removing candidate and confirmed binaries significantly cleans up the color-period distribution in Figures~\ref{fig:cpdnew} and \ref{fig:cpdtt}, revealing very clearly the well-defined slow-rotating sequence that extends from F3 to $\approx$M2 stars in the cluster.

\begin{figure*}
    \centering
    \includegraphics[width=\textwidth,keepaspectratio]{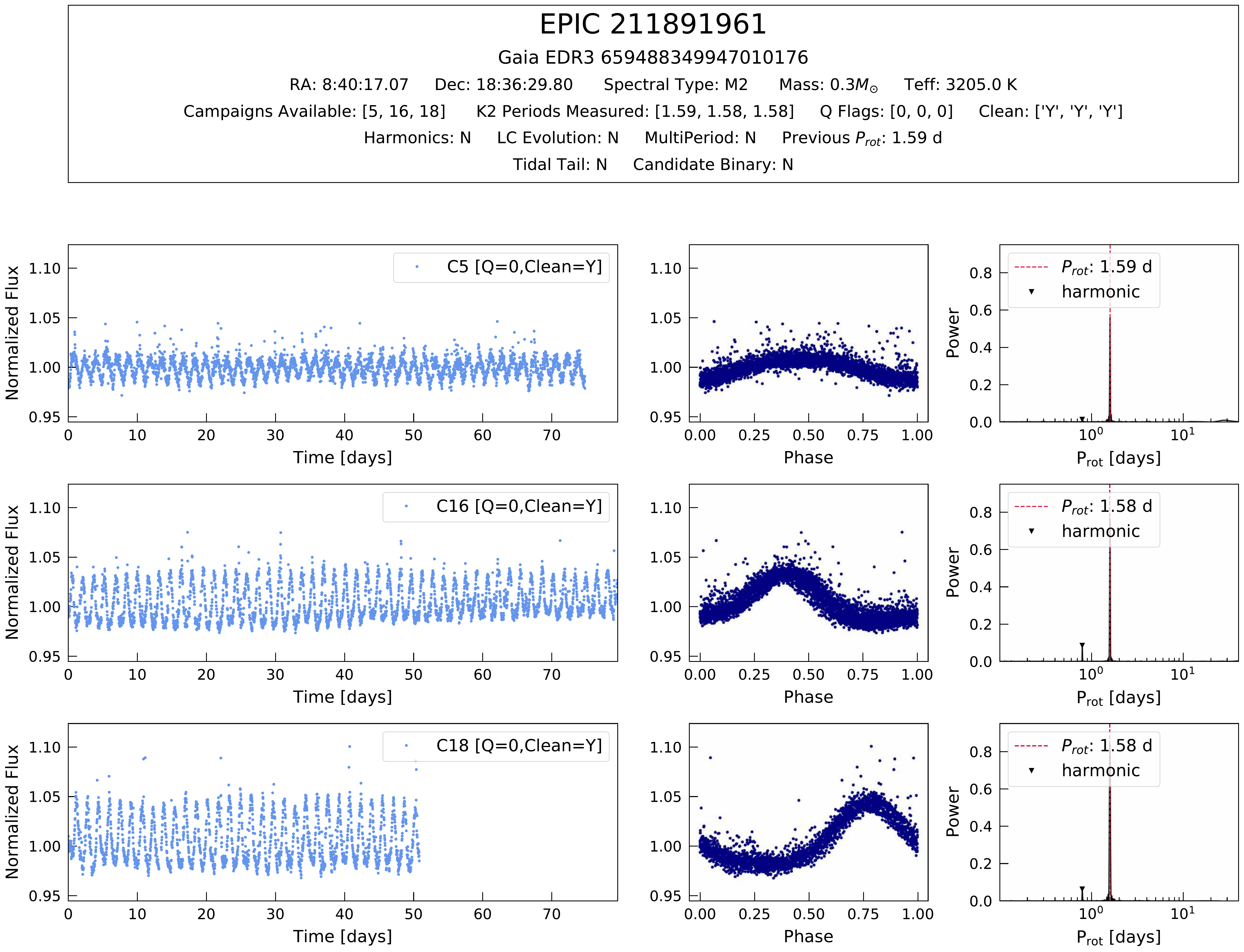}
    \caption{Example figure set for a star with multiple \ktwo\ light curves and \prot\ measured, EPIC 211891961. \textit{Top:} Information about the star including the EPIC ID, \Gaia\ EDR3 name, right ascension, declination, mass, effective temperature, campaigns observed, periods measured, quality flags assigned, clean detection flag, harmonics detected across campaigns, light-curve evolution observed, multiple periods observed, previous \prot\ measured, tidal-tail  status, and binary status. \textit{Left column:} Light curves from C5, C16, and C18. \textit{Middle column:} Light curves phase-folded to \prot\ measured from each campaign. \textit{Right column:} LS periodograms. The complete figure set is available in the online journal.} 
    \label{fig:lccomp}
\end{figure*}

For the 825 stars with multiple campaigns' worth of data, we also provide a figure showing the different light curves and corresponding LS periodograms, along with information about the stars (see Figure~\ref{fig:lccomp} for an example).

When comparing our \Prot\ for the 793 stars with a previously measured \Prot, we find that 752 ($>$94\%) of them are in agreement to within 10\% (median difference: 0.5\%, standard deviation: 2\%). This supports the validity of our methods and \prot\ measurement. Of the 41 stars that have discrepant \Prot, in 19 cases one of the reported \Prot\ is a harmonic, and in six cases multiple periods are reported for the star. The remaining 16 stars with discrepant \prot\ had  periods measured from ground-based observations or solely from C5 observations (either in D17 or R17). Here, we adopt our \ktwo\ period given the improvement in light curve quality compared to ground-based observations and repeated observations compared to the \prot\ measured solely from C5. Of the 793 \prot\ we recover, there is no case where the differences between our \prot\ and the literature measurement are due to significant astrophysical reasons.

\begin{deluxetable*}{lllcl}
\tablecaption{Description of Lomb-Scargle outputs for the Praesepe rotators \label{t:ls}}
\tablewidth{0pt}
\tablehead{
\colhead{Column} & \colhead{Format} & \colhead{Units} & \colhead{Example} & \colhead{Description}
}
\startdata
\sidehead{\textit{Identifiers:}} 
EPIC    & integer & \nodata & 211891961 & \textit{K2} EPIC ID \\
EDR3Name & string & \nodata & Gaia EDR3 659488349947010176 & \textit{Gaia} EDR3 Source ID \\
2MASS & string & \nodata & 2MASS J08401707+1836298 & 2MASS Source ID \\
\sidehead{\textit{Coordinates:}} 
RA      & float & degrees & 130.07095 & Right ascension \\
DEC     & float & degrees & 18.60823 & Declination \\
\sidehead{\textit{\prot\ Data:}} 
Prot    & float  & days    & 1.58 & \prot\ chosen for this star \\
\\
C5Prot    & float  & days    & 1.59 & Primary \prot\ measured in C5  \\
C5Power     & float  & \nodata    & 0.56 & LS power for C5Prot  \\
C5Sprot     & float  & days    & 0.79 & Secondary \prot\ measured in C5 \\
C5Spower     & float  & \nodata     & 0.01 & LS power for C5Sprot \\
C5Clean    & integer  & \nodata     & 1 & Clean detection for C5 LC (C5Sprot/C5prot < 0.6)? Y=1, N=0 \\
C5Q    & integer  & \nodata & 0 & Assigned quality flag for C5 LC \\
C5Rvar    & float  & mag & 0.029 & R\textsubscript{var} for C5 LC\\ 
C5thresh    & float  & \nodata & 0.009 & Minimum LS power needed for C5 \prot\ detection  \\
C5SNR    & integer  & \nodata & 38 & C5 lightcurve SNR  \\
\\
C16Prot & float  & days     & 1.58 & Primary \prot\ measured in C16 \\
C16Power    & float  & \nodata  & 0.77   & LS power for C16Prot \\
C16Sprot    & float  & days     & \nodata & Secondary \prot\ measured in C16 \\
C16Spower      & float   & \nodata     & \nodata & LS power for C16Sprot \\
C16Clean    & integer   & \nodata & 1  & Clean detection for C16 LC (C16Sprot/C16prot < 0.6)? Y=1, N=0 \\
C16Q    & integer  & \nodata     & 0 & Assigned quality flag for C16 LC \\
C16Rvar    & float  & mag     & 0.053 & R\textsubscript{var} for C16 LC \\
%
%
C16thresh     & float   & \nodata  & 0.008   & Minimum LS power needed for C16 \prot\ detection  \\
C16SNR    & integer  & \nodata & 38 &  C16 lightcurve SNR \\
\\
C18Prot    & float  & days    & 1.58 & Primary \prot\ measured in C18 \\
C18Power     & float  & \nodata    & 0.84 & LS power for C18Prot \\
C18Sprot     & float  & days    & 0.79 & Secondary \prot\ measured in C18 \\
C18Spower     & float  & \nodata     & 0.06 & LS power for C18Sprot \\
C18Clean    & integer  & \nodata     & 1 & Clean detection for C18 LC (C18Sprot/C18prot < 0.6)? Y=1, N=0 \\
C18Q    & integer  & \nodata & 0 & Assigned quality flag for C18 LC \\
C18Rvar    & float  & mag & 0.068 & R\textsubscript{var} for C18 LC\\ 
C18thresh    & float  & \nodata & 0.01 & Minimum LS power needed for C18 \prot\ detection \\
C18SNR   & integer  & \nodata & 37 & C18 lightcurve SNR  \\
\sidehead{\textit{Additional Info:}} 
QFClean  & integer  & \nodata    & 1     & $\geq$1 high-quality flag (Q = 0) for star's lightcurve(s)? Y=1, N=0 \\
MultiPeriod    & integer   & \nodata      & 0  & Multiple periods found in light curve(s)? Y=1, N=0 \\      
Neighbors      & integer   & \nodata & 0 & Possible contamination from neighboring star in aperture? Y=1, N=0 \\
Harmonics     & integer   & \nodata    & 0   & Strong harmonics found within and/or across lightcurve(s)? Y=1, N=0  \\
LCEvolution    & integer   & \nodata     & 0   & LC evolution found in light curve(s)? Y=1, N=0  \\
PreviousProt     & float & days & 1.59     & Previous \prot\ for star \\ 
Binary     & integer & \nodata  & 0     & Binary flag \\ 
%
%
%
%
%
%
\enddata
\end{deluxetable*}

\begin{deluxetable*}{lllcl}
\tablecaption{Description of Rotators in Praesepe \label{t:prot}}
\tablewidth{0pt}
\tablehead{
\colhead{Column} & \colhead{Format} & \colhead{Units} & \colhead{Example} & \colhead{Description}
}
\startdata
\sidehead{\textit{Identifiers:}} 
EPIC    & integer & \nodata & 211891961 & \textit{K2} EPIC ID \\
EDR3Name & string & \nodata & Gaia EDR3 659488349947010176 & \textit{Gaia} EDR3 Source ID \\
2MASS & string & \nodata & 2MASS J08401707+1836298 & 2MASS Source ID \\
%
\sidehead{\textit{Gaia EDR3 data:}} 
RA      & float & degrees & 130.07095 & Right ascension \\
DEC     & float & degrees & 18.60823 & Declination \\
pmra    & float  & mas    & $-36.711$ & Right ascension proper motion \\
pmdec    & float  & mas    & $-11.973$ & Declination proper motion \\
e\_pmra     & float  & mas    & 0.12 & Proper motion in RA error \\
e\_pmdec     & float  & mas    & 0.079 & Proper motion in DEC error \\
plx     & float  & mas     & 5.553 & Parallax \\
eplx    & float  & mas     & 0.132 & Parallax error \\
D    & float  & pc     & 178.313 & Distance from \cite{edr3dists} \\
eD    & float  & pc     & 4.217 & Distance error \\
epsi    & float  & \nodata & 0.168 & Astrometric excess noise \\
sepsi    & float  & \nodata & 0.618 & Significance of astrometric excess noise \\ 
ruwe    & float  & \nodata & 1.045 & Re-normalised Unit-Weight Error \\
bp\_rp  & float  & mag     & 2.92 & \textit{Gaia} EDR3 color: G\textsubscript{BP} $-$ G\textsubscript{RP} \\
Gmag    & float  & mag     & 17.27 & \textit{Gaia} EDR3 $G$ magnitude \\
Kmag    & float  & mag     & 13.35 & 2MASS $K$ magnitude \\
RV      & float   & \kms     & \nodata & Radial Velocity \\
e\_RV     & float   & \kms & \nodata   & RV error  \\
IncompleteGaiaFlag    & integer  & \nodata     & 0 & Flag if missing \textit{Gaia} EDR3 measurements \\
%
\sidehead{\textit{Binarity indicators:}} 
%
dPM     & float   & \mas     & 1.19   & Proper motion deviation from cluster \\ 
dCMD\_bprp    & float   & mag      & $-0.035$  & Photometric excess in  M\textsubscript{G} vs. G $-$ G\textsubscript{RP} \\      
dCMD\_grp       & float   & mag     & $-0.069$ & Photometric excess in M\textsubscript{G} vs. G\textsubscript{BP} $-$ G\textsubscript{RP} \\
dRV     & float   & \kms     & \nodata   & RV Deviation from cluster  \\
MultiPeriod     & integer   & \nodata    & 0   & Multiple periods found in light curve? Y = 1, N = 0  \\
AstrometricBinary     & integer & \nodata  & 0     & Astrometric binary flag (dPM $\geq 2.5$ mas) \\ 
Photometric Binary     & integer & \nodata  & 0     & Photometric binary flag (either dCMD $\geq 0.375$) \\ 
LiteratureBinary     & integer & \nodata  & 0     & Literature binary flag \\ 
WideBinary     & integer & \nodata  & 0     & Wide binary flag (ruwe $\geq 1.2$) \\ 
Binary     & integer & \nodata  & 0     & Binary flag \\ 
%
\sidehead{\textit{Additional Information:}} 
TidalTail     & integer & \nodata   & 0 & Member of tidal tails? Y=1, N=0 \\
k2data      & integer & \nodata   & 1 & Star observed by \ktwo? Y=1, N=0  \\
%
\sidehead{\textit{Stellar Properties:}} 
Teff   & integer & K       & 3205 & Effective temperature \\
Mass   & float   & \msun   & 0.3 & Mass \\
SpT    & string  & \nodata & M4 & Spectral type \\
\sidehead{\textit{Rotation data:}} 
%
Prot     & float  & days    & 1.58   & Rotation period measured in this work \\
QFClean  & integer  & \nodata    & 1     & $\geq$1 high-quality flag for star's lightcurve(s)? Y = 1, N = 0 \\
PreviousProt      & float  &  days       & 1.59    & Previous \prot\ for star \\
LCEvolution    & integer  & \nodata    & 0 & Morphology evolution in \ktwo\ light curve(s)?  Y = 1, N = 0 \\
\enddata
\end{deluxetable*}

\section{Discussion}\label{section:discussion}

The addition of 220 stars to the sample of Praesepe rotators does not significantly change the color-period distribution for the cluster (see Figure~\ref{fig:cpdnew}). The bluest stars in the sample are rapidly rotating; the turnover to slower \prot\ occurs roughly at G\textsubscript{BP} $-$ G\textsubscript{RP} = 0.9. From there, as the stars decrease in mass, their \Prot\ increases. This is thought to be because the efficiency of their angular-momentum loss increases as their convection zones extend deeper from the surface of the star \citep{barnes2010}. 
After spectral type M3, we continue to see the typical sharp transition to rapid rotation for fully convective stars, 
which have not yet begun to spin down at Praesepe's age.

Most of the tidal-tail stars recently identified by \cite{Roser19} fall outside of the C5, 16, and 18 fields-of-view. Of the 318 stars in these tails, we measure a \prot\ for 33. Twenty-three of these are candidate single rotators. We show the overall \prot\ distribution for these stars and then isolate the candidate single rotators in Figure~\ref{fig:cpdtt} (these stars are also included in the new \Prot\ sample shown in Figure \ref{fig:cpdnew}). 

\begin{deluxetable*}{ccccccc}[bh!]
\centering 
\tabletypesize{\footnotesize} 
\tablewidth{0pt}
\tablecaption{Single-star rotators observed by \ktwo}
\label{tbl:highqfoverlap}
\tablehead{
\colhead{Campaign} & 
\colhead{Praesepe Members} &
\colhead{\Prot\ measured} & 
\colhead{Single with Q = 0} &
\colhead{Also in C5} & 
\colhead{Also in C16} &
\colhead{Also in C18} 
}
\startdata
5 & 911 &	850 & 398 & \nodata & 120 & 308  \\ 
16 & 512 & 466 & 211 & 120 & \nodata & 127  \\
18 & 894 & 831 & 350 & 308 & 127 & \nodata
\enddata
\tablecomments{We measure a \prot\ for 311 of the 327 Praesepe members observed by \ktwo\  had \prot\ measured in all three campaigns. Of these, 112 were likely single stars with Q = 0.}
\end{deluxetable*}

The \prot\ distribution for tidal-tail stars follow the \prot\ distribution of the cluster core. 
This provides additional evidence that that these stars are indeed cluster members. Random field stars are likely to be much older than those in Praesepe, and we would expect to see much smaller photometric modulations and/or longer \prot\ if these stars were not members of the cluster. 

\prot\ measurements can therefore offer additional evidence for membership for stars located several tidal radii away from a cluster's core. This could be particularly valuable when searching for planets in clusters in general and with {\it TESS} specifically. Not only are there more stars associated with a given cluster to search, but targeting stars away from the crowded cluster core could also ease issues related to blending, a known problem for \textit{TESS}. Additionally, stars in the tidal tails of clusters are likely less susceptible to dynamical processes (e.g., tidal interactions) that occur in the cores of open clusters and disrupt planetary systems (e.g., \citealt{fujii19}). Thus, planet occurrence rates may be higher in the tidal tails.  


\subsection{Light-Curve Evolution}\label{section:spotev}
Evolving spot configurations, due to differential rotation, to dynamo cycles, or to the growth and decay of individual active regions, can cause phase drifts or evolving patterns in a light curve. \textit{Kepler} light curves showed evidence for these effects in older field stars \citep{vida2014,davenport2015,reinhold2015}, and \textit{K2}'s repeat observations of Praesepe provide the first opportunity to potentially study similar effects in a middle-aged open cluster. 

\begin{figure*}
    \centering
    \includegraphics[width=\textwidth,keepaspectratio]{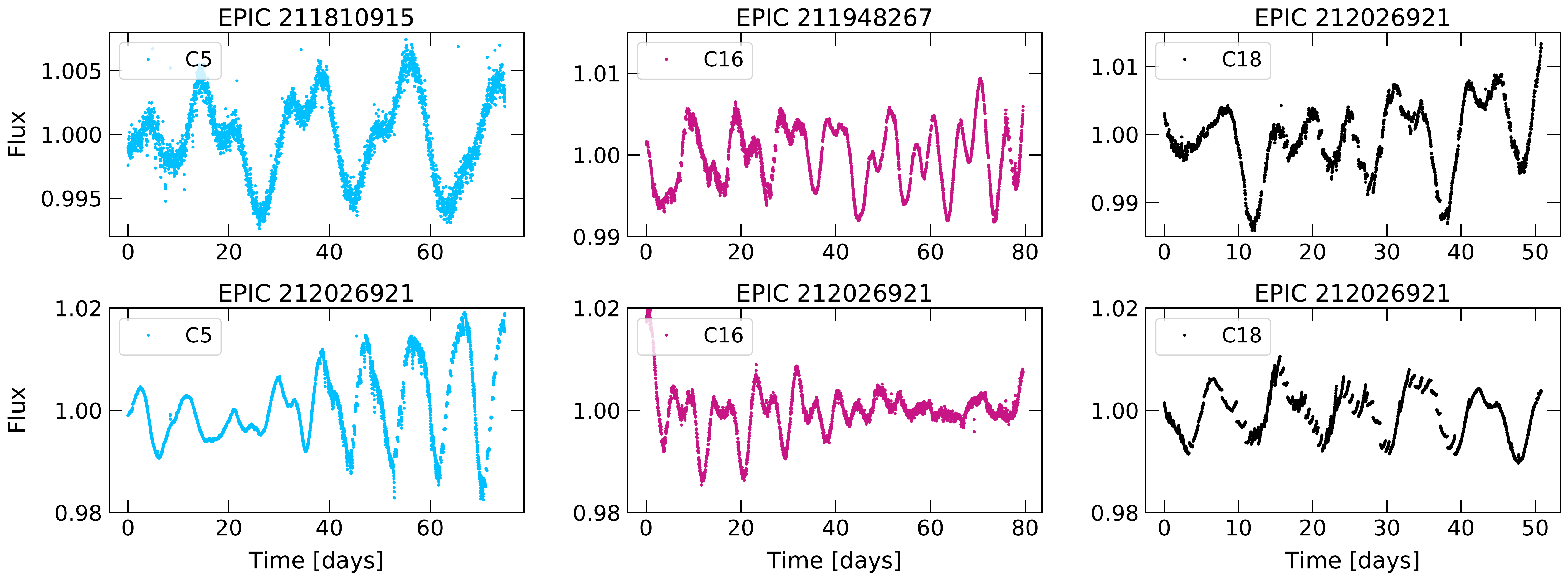}
    \caption{Examples of light-curve evolution in various rotators within a single campaign's observations (top three panels) and of a single rotator's light-curve evolution across campaign observations (bottom row).}
    \label{fig:spev}
\end{figure*}

As mentioned in Section \ref{section:methods}, we flag stars for light-curve evolution through by-eye examinations of every star's light curves. Of the 1013 rotators, we find that 386 (38\%) have visible evolution within a campaign and/or across campaigns. Examples are provided in Figure~\ref{fig:spev}.

\begin{figure}
    \centering
    \includegraphics[scale=0.49]{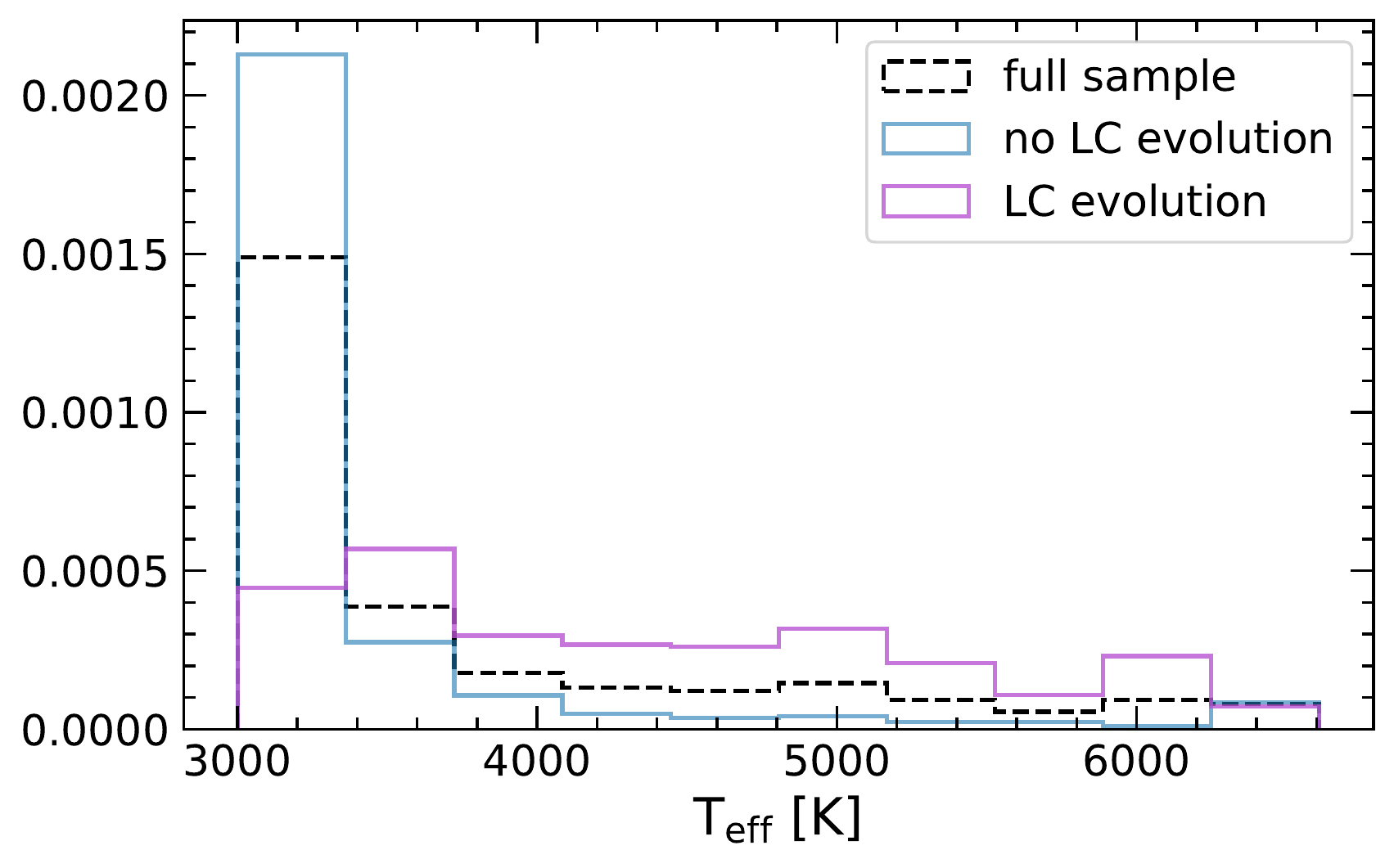}
    \caption{Normalized effective temperature distributions for the full set of rotators (grey dashed histogram), rotators flagged for light-curve evolution (pink), and rotators showing no major light-curve evolution (blue). Light curves flagged for morphological evolution are more likely to be those for hotter (higher mass) stars.}
    \label{fig:spevhist}
\end{figure}

We find that the light curves for higher-mass and hotter-temperature ($\gtrsim$3400 K) stars show more morphological changes during \textit{K2} observations than those for low-mass, cooler stars (see Figure~\ref{fig:spevhist}). 

Similar results were found for \textit{Kepler} stars and for members of the Pleiades and Blanco 1 open clusters \citep{reinhold13,stauffer16,gillen20}. 

\indent One potential explanation for these morphological changes in the light curves is that they are a result of differential rotation in the star. One might expect that it is easier to observe differential rotation in a higher mass star, where the effect is more obvious. 
Another possibility is that these changes are due to evolving spot patterns on the star. Spots on the Sun have lifetimes of days to a few months, with most spots decaying in less than one rotation period \citep{petrovay1997}. 

Disentangling which of these mechanisms is causing these changes in the light curves is difficult, especially if they are occurring on the same time scale \citep{2020basri}. 
Modeling these light curves with a Gaussian Process model (e.g., using \textit{starry}; \citealt{starry}) may lend insight into the spot evolution that is occurring and provide a metric that can quantitatively measure such changes in the light curve (e.g., \citealt{gordon2020}). Investigating this is beyond the scope of this work, but we note this as a point of  future exploration.

\subsection{Stability of \prot\ Measurements Between Campaigns}

\begin{figure*}
    \centering
    \includegraphics[width=\textwidth,keepaspectratio]{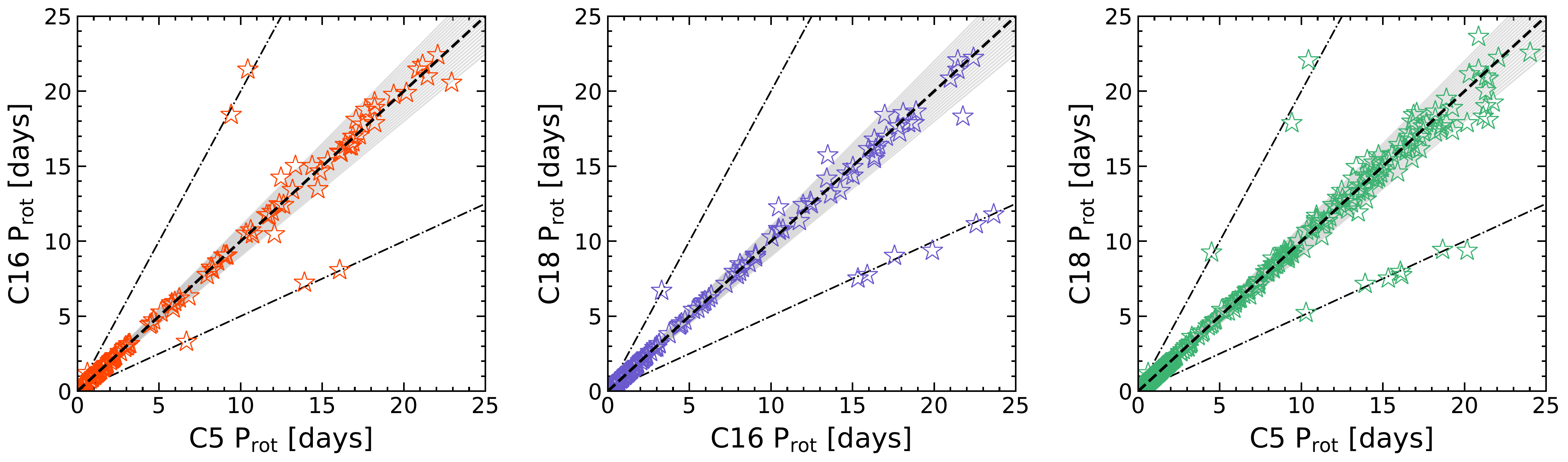}
    \caption{Comparison between \Prot\ measured for single-star rotators with multiple high-quality \ktwo\ light curves. The dot-dashed lines on either side of the dashed, 1:1 line are the 1:2 and 2:1 lines corresponding to half- and double-period harmonics. The grey band corresponds to a difference $\leq$10\% from a 1:1 match. \textit{Left:} C5 vs.~C16 \prot. \textit{Middle:} C16 vs.~C18 \prot. \textit{Right:} C5 vs.~C18 \prot. The vast majority of stars have \Prot\ that differ by less than 10\% from one campaign to another.} 
    \label{fig:perpercomp}
\end{figure*}

Because \ktwo\ observed Praesepe three different times over a $\approx$3 year baseline, we can test our ability to recover \Prot\ measurements. 
We compare \prot\ detections across campaigns for single stars that have $Q = 0$ (see Figure \ref{fig:perpercomp}).
With this sample of 331 stars (see  Table~\ref{tbl:highqfoverlap}), we find that in $>$95\% of the cases, the measured periods agree to within 10\%, with a median difference of 0.3\% and standard deviation of 2\%, and conclude that we are measuring the star's intrinsic \Prot\ rather than a chance spot alignment.

By comparison, \cite{reinhold20} found that, when measuring rotation periods for all of the stars with \ktwo\ light curves (i.e., from Campaigns 0 to 18), 75-90\% of the stars with multiple observations had \prot\ measurements within 20\% from different campaigns. 
We attribute the higher precision of our results to surveying a specific, single-aged population of stars, and to using a robust combination of visual and automated inspection to identify high-quality light curves/\Prot\ detections. Additionally, the \cite{reinhold20} sample includes many field stars, which on average have intrinsically longer \prot\ than Praesepe rotators and adds measurement uncertainty as compared to our work.

We also find that we can measure the same \Prot\ to better than 10\% in spite of visually identified spot evolution and/or differential rotation, as is the case with 103 of the 331 stars in our sample of rotators with multiple \ktwo\ light curves. Just as with the entire \prot\ catalog, the correlation between higher temperatures and light-curve evolution seen in Figure \ref{fig:spevhist} persists for these 331 stars. 

This test of \prot\ stability provides some insight into what typical uncertainties one might place on \Prot\ measurements.

Because we find that the \Prot\ are recovered to within 10\% at such a high rate, we argue that this is the right typical uncertainty for \Prot\ found using LS periodograms. 

How this uncertainty compares to that returned by other \Prot\ measurement techniques, such as the Auto-Correlation Function (ACF), Gaussian processes, or machine learning (e.g., \citealt{mcquillan2014,angus2018,Lu2020}),  is still unclear. 
Each of these techniques may be better at measuring robust \Prot\ in specific situations. For example, \cite{curtisrup14720} finds that the ACF performs better than LS for measuring \Prot\ in older stars that tend to have double-dip signals \citep{basri18}. We discuss this further in Section \ref{sec:discrepant}.

\subsection{Stars with Half- or Double-Period Harmonic \Prot\ Measurements}
 
As mentioned in Section~\ref{section:finalprot}, we measure the double- or half-period harmonic \prot\ for 234 stars, or $\approx$23\% of our rotator sample of 1013 stars. From the sample of likely-single stars with multiple observations of Q = 0,  these measurements of \Prot\ harmonics occur at a much lower rate of $\approx$3-6\%. When comparing the 262 stars that previously had ground-based \prot\ 
to those measured from \ktwo, we find that only eight stars have harmonic \Prot\ measurements.

\indent The frequency at which harmonics of the true period are measured naturally increases if the length of the observations differs by a large amount, as was the case with C18 compared to C5 and C16 (see increased number of stars with half- and double-period harmonics in right two panels of  Figure \ref{fig:perpercomp}). This has important implications for current and future photometric missions. More care must be taken in validating a \prot\ if there is only one observation, particularly if it has a short baseline (e.g., a star observed during a single \textit{TESS} sector). 

\begin{figure}
    \centering
    \includegraphics[scale=0.45]{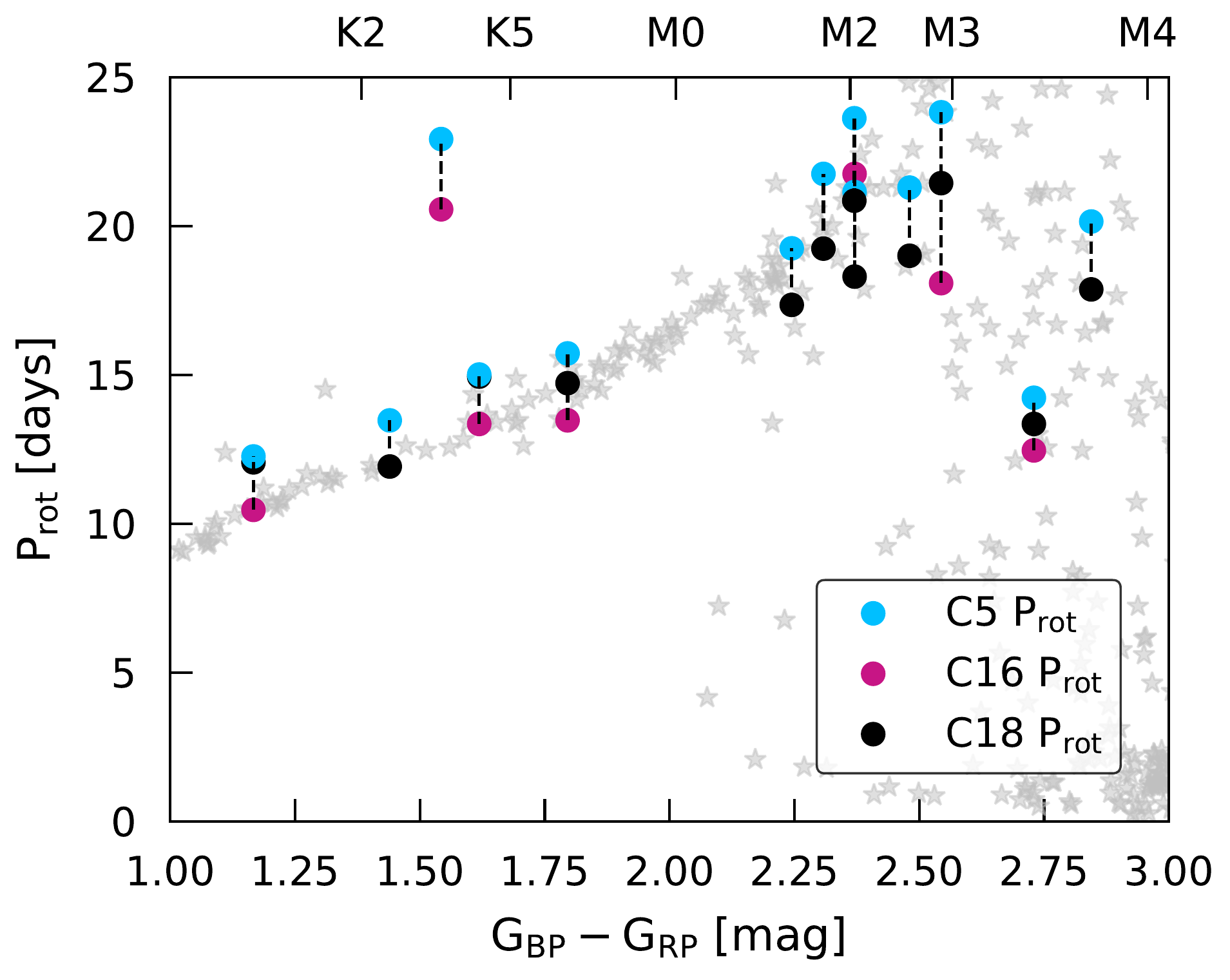}
    \caption{\Gaia\ color-period distribution of single stars with  \prot\ discrepant by more than 10\% between campaigns. \prot\ measurements for each star denoted with light blue circle for C5, magenta circle for C16, and black circle for C18, connected by black dashed line. The stable \prot\ single-rotator sample is shown as the light grey stars.}
    \label{fig:discrepant}
\end{figure}

\begin{figure*}
    \centering
    \includegraphics[height=\textheight,width=\textwidth,
    keepaspectratio]{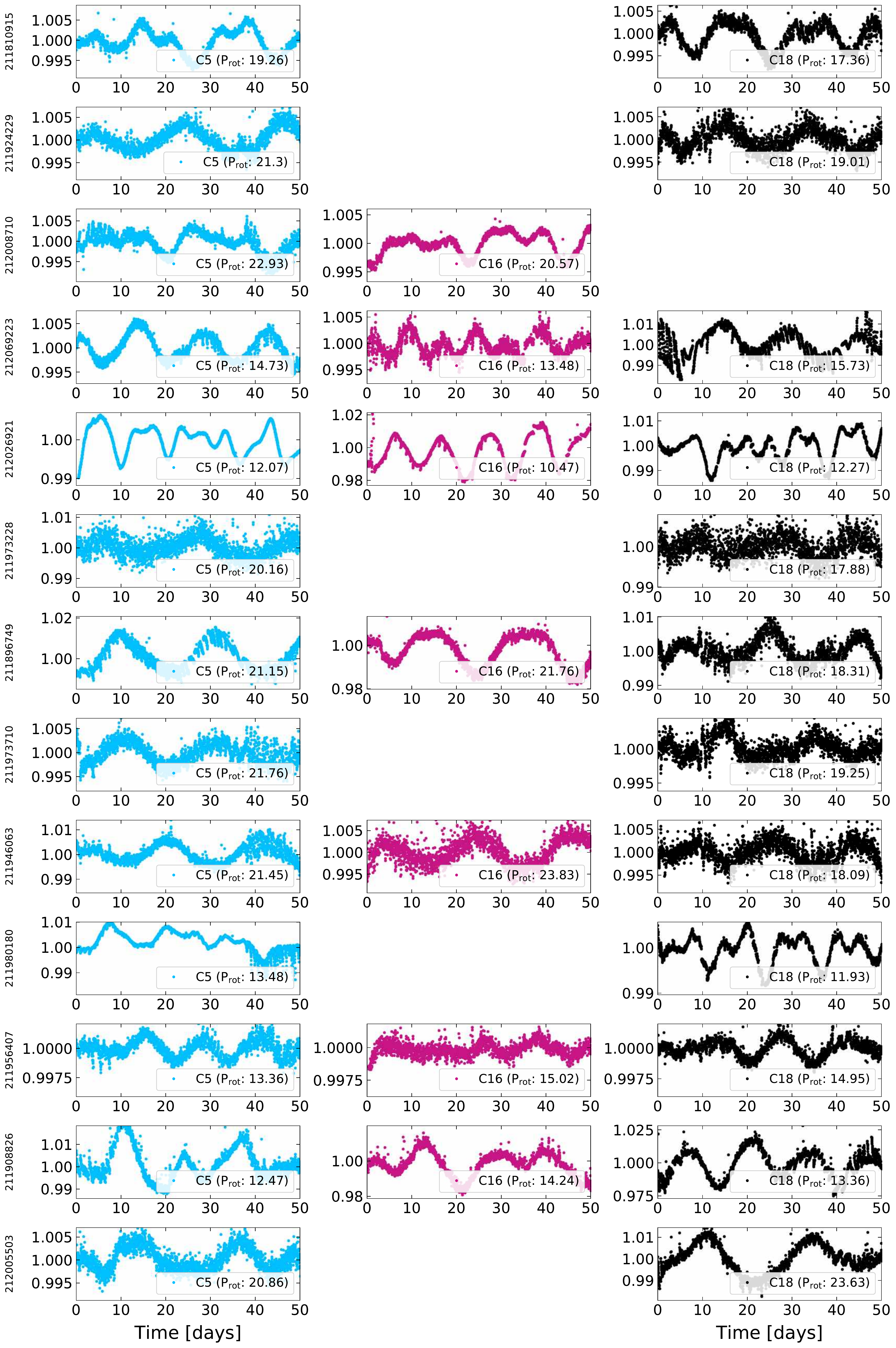}
    \caption{Light curves of rotators with discrepant \prot\ measurements between campaigns. \textit{Left}: C5 light curves. \textit{Middle}: C16 light curves. \textit{Right}: C18 light curves. Additional analysis and/or processing of 10 of these light curves resolves the discrepancy between the \prot\ measured in different campaigns. The three exceptions are EPICs 211924229, 211896749, and 211946063.}
    \label{fig:dlcs}
\end{figure*}

\subsection{Stars with Discrepant \prot\ Measurements}\label{sec:discrepant}

Only 13 stars in our sample have \Prot\ discrepant by more than 10\% (see Table \ref{tbl:drot}), and the discrepancy is only 11-18\%. These stars are distributed throughout the cluster's color-period plane, as shown in Figure~\ref{fig:discrepant}, so there is no obvious dependence on color. Ten of these stars have discrepant periods from C18. Since this campaign was the shortest of the three, there is likely a bias toward finding shorter periods when working with its light curves, which is more likely to result in a larger discrepancy with previously measured \Prot\ for slower rotators. This is largely confirmed by Figure~\ref{fig:discrepant}, where the C18 \Prot\ is generally the shortest measured.

The four stars that have \Prot\ discrepant by more than 10\% between C5 and C16 (one of which also has a discrepant measurement in C18) do show a significant amount of morphological evolution in their light curves. This could be the result of differential rotation and/or spot evolution that is strong enough to affect the measurement of \prot. 

We also note that light curves for a majority of the 13 stars show double-dipping patterns. These are not well-suited to a LS analysis, as the latter assumes a sinusoidal signal. We use the ACF on all of 13 and find that this results in 10 cases in \Prot\ measurements that differ by $<$10\%. This shows that pattern matching techniques like the ACF are probably better suited for measuring \prot\ for double-dipping stars.\footnote{While the ACF might not be ideal for \textit{TESS}, there are alternatives such phase dispersion minimization (\citealt{Barnes2016}), which can be applied to unevenly sampled time series.}
 
Light curves for six of the 13 stars feature instrumental systematics, presumably introduced by \ktwo's rolling, which persist despite the PDCSAP corrections.
We apply pixel-level decorrelation (PLD, \citealt{Deming_2015}) to the \ktwo\ target pixel files for these six stars. This identifies trends in the pixels surrounding the star, which ideally model the telescope's rolling motion and can be subtracted from the target's light curve. 
In several cases, this also allows us to measure \prot\ that are within 10\% of these stars' other \prot\ measurements.

We include all of the discrepant \prot\ stars' light curves in Figure \ref{fig:dlcs}.

\begin{deluxetable*}{lccccccccccc}
\centering 
\tabletypesize{\footnotesize} 
\tablecaption{Rotators with discrepant \prot\ measurements between \ktwo\ campaigns}
\label{tbl:drot}
\tablehead{
\colhead{EPIC} & 
\colhead{C5\Prot} & 
\colhead{C5QF} &
\colhead{C16\Prot} &
\colhead{C16QF} &
\colhead{C18\Prot} &
\colhead{C18QF} &
\colhead{Discrepant Campaigns} &
\colhead{LC Evolution} &
\colhead{SpT} &
\colhead{T$_{eff}$ (K)} &
}

\startdata
211810915 & 19.26  & 0    & \nodata & \nodata & 17.36   & 0     & C5/C18              & 1           & M2   & 3555 \\
211924229 & 21.3   & 0    & 0.25    & 3       & 19.01   & 0     & C5/C18              & 1           & M3   & 3418 \\
212008710 & 22.93  & 0    & 20.57   & 0       & 11.26   & 1     & C5/C16              & 1           & K5.5 & 4284 \\
212069223 & 14.73  & 0    & 13.48   & 0       & 15.73   & 0     & C16/C18             & 0           & K9   & 3921 \\
212026921 & 12.07  & 0    & 10.47   & 0       & 12.27   & 0     & C5/C16, C16/C18     & 1           & K2.5 & 4992 \\
211973228 & 20.16  & 0    & \nodata & \nodata & 17.88   & 0     & C5/C18              & 0           & M4   & 3221 \\
211896749 & 21.15  & 0    & 21.76   & 0       & 18.31   & 0     & C5/C18, C16/C18     & 0           & M2.5 & 3507 \\
211973710 & 21.76  & 0    & \nodata & \nodata & 19.25   & 0     & C5/C18              & 0           & M2   & 3529 \\
211946063 & 21.45  & 0    & 23.83   & 1       & 18.09   & 0     & C5/C18              & 0           & M3   & 3371 \\
211980180 & 13.48  & 0    & \nodata & \nodata & 11.93   & 0     & C5/C18              & 1           & K5   & 4428 \\
211956407 & 13.36  & 0    & 15.02   & 0       & 14.95   & 0     & C5/C16,C5/C18       & 1           & K6.5 & 4204 \\
211908826 & 12.47  & 0    & 14.24   & 0       & 13.36   & 0     & C5/C16              & 1           & M3.5 & 3246 \\
212005503 & 20.86  & 0    & \nodata & \nodata & 23.63   & 0     & C5/C18              & 1           & M2.5 & 3507                     
\enddata
\tablecomments{Only \Prot\ measurements with an assigned quality flag (QF) = 0 were considered for the \prot\ comparison analysis.}
\end{deluxetable*}

\section{Conclusions}\label{section:conclusion}

We present the most up-to-date and complete sample of rotators for the $\approx$670-Myr-old open cluster Praesepe, as measured from \ktwo\ light curves, and provide \Gaia\ EDR3 astrometry for these stars. As \ktwo\ observed  Praesepe three times (in its Campaigns 5, 16, and 18) over $\approx$3 year baseline, we also test the \prot\ stability from campaign to campaign for stars observed more than once. 

\begin{itemize}
    \item \textit{K2} yielded 1013 rotation periods, 465 of which are for stars we identify as candidate or confirmed binaries. Our results are summarized in Figure \ref{fig:cpdnew}. The identification of candidate and confirmed binaries significantly reduces the scatter in the color-period distribution, particularly for stars on the slow-rotating sequence that extends from F3 to M2 stars in the cluster. In addition to these stars, there are 17 rotators that do not have \ktwo\ data and were discovered using earlier ground-based observations. In Table \ref{t:prot}, we present the entire catalog of 1030 Praesepe rotators, a total that corresponds to 63\% of the stars with masses $\lapprox 1.3$~\Msun\ in our membership catalog for the cluster.
    \item We measure new \prot\ for 220 Praesepe stars and recover \prot\ measurements for 793 of the 812 rotators previously reported in the literature. Seventeen of the 19 stars for which we did not recover a \prot\ did not have \ktwo\ observations, and the \prot\ we measure for the remaining two stars' from their \ktwo\ light curves do not pass our validation processes. 
    \item Adding these 220 new rotators to the existing \Prot\ catalog does not significantly change the distribution of Praesepe stars in color-period space. 
    \item These 220 new \Prot\ include \Prot\ for 33 stars recently identified as belonging to Praesepe's tidal tails \citep{Roser19}. These 33 stars follow the cluster's overall color-period distribution (see Figure~\ref{fig:cpdtt}), strengthening their association with Praesepe.
    \item For 38\% of the rotators, we observe morphological evolution within a single light curve and/or across campaign light curves. We find that this occurs more often for the higher-mass stars in our sample. This is likely a result of spot evolution and/or differential rotation.
    \item Of the 331 stars with multiple high-quality observations, we measure the same \prot\ to better than 10\% in $>$95\% of the cases. This suggests that the uncertainty on any individual \Prot\ measurement in our catalog is better than 10\%; the median difference in the \Prot\ measurements is 0.3\%, with and standard deviation of 2\%.  
    \item Ten of the 13 stars with \Prot\ measurements discrepant by more than 10\% were observed in C18. C18 was significantly shorter than the other two campaigns, which could result in an underestimate of longer \Prot\ and the observed discrepancies. 
    We also note that using methods like the ACF and PLD to measure \prot\ or process light curves for these stars results in smaller differences in the \prot\ relative to that measured in other campaigns.  
    In all 13 cases, the \Prot\ obtained from different catalogs are still within $\leq$18\% of each other. 
    
\end{itemize}

This work is a first step to understanding the impact of spot evolution and of differential rotation on large-scale surveys of \Prot. Future work  includes Gaussian-process modeling of these light curves to simulate spot evolution, and simultaneous spectroscopic and photometric observations to deepen our understanding of the connection between rotation and activity. Additionally, combining past ground-based observations with \ktwo\ observations, as well as with data from {\it TESS} observations of Praesepe, will produce a sample of stars with light curves collected over a $\gapprox$10 year baseline. With these data, we will continue to develop our understanding of magnetic cycles and the rotation-activity relation for low-mass stars. 

\acknowledgments
We thank the anonymous referee for their very positive review. We thank Ruth Angus and participants in the THYME Workshop for helpful discussions and advice. 

R.R.~gratefully acknowledges the support of the Columbia University Bridge to the Ph.D.~Program in STEM. M.A.A.~acknowledges support provided by the NSF through grant AST--2009840 and by NASA through grants 80NSSC18K0448, 80NSSC19K0383, and 80NSSC19K0636. 

This paper includes data collected by the \Kepler\ mission. Funding for the \Kepler\ mission is provided by the NASA Science Mission directorate. Some of the data presented in this paper were obtained from the Mikulski Archive for Space Telescopes (MAST). STScI is operated by the Association of Universities for Research in Astronomy, Inc., under NASA contract NAS5--26555. Support for MAST for non--HST data is provided by the NASA Office of Space Science via grant NNX13AC07G and by other grants and contracts. This work has made use of data from the European Space Agency (ESA) mission {\it Gaia}\footnote{\url{https://www.cosmos.esa.int/gaia}}, processed by the {\it Gaia} Data Processing and Analysis Consortium (DPAC)\footnote{\url{https://www.cosmos.esa.int/web/gaia/dpac/consortium}}. Funding for the DPAC has been provided by national institutions, in particular the institutions participating in the {\it Gaia} Multilateral Agreement.


%

\vspace{5mm}
\facilities{\Kepler/\textit{K2}, \Gaia, MAST, CDS, ADS}


\software{astropy \citep{astropy:2013,astropy18}, lightkurve \citep{lightkurve}, scipy \citep{scipy}, K2SC \citep{aigrain2016},
astroML \citep{astroML}, astroquery \citep{astroquery}, k2fov \citep{mullally2016}, starspot }

\appendix 
\renewcommand\thefigure{\thesection.\arabic{figure}}

\section{Quality Flag Designation}\label{sec:qfd}
\setcounter{figure}{0}
\begin{figure*}[htp]
    \centering
    \includegraphics[width=\textwidth,keepaspectratio]{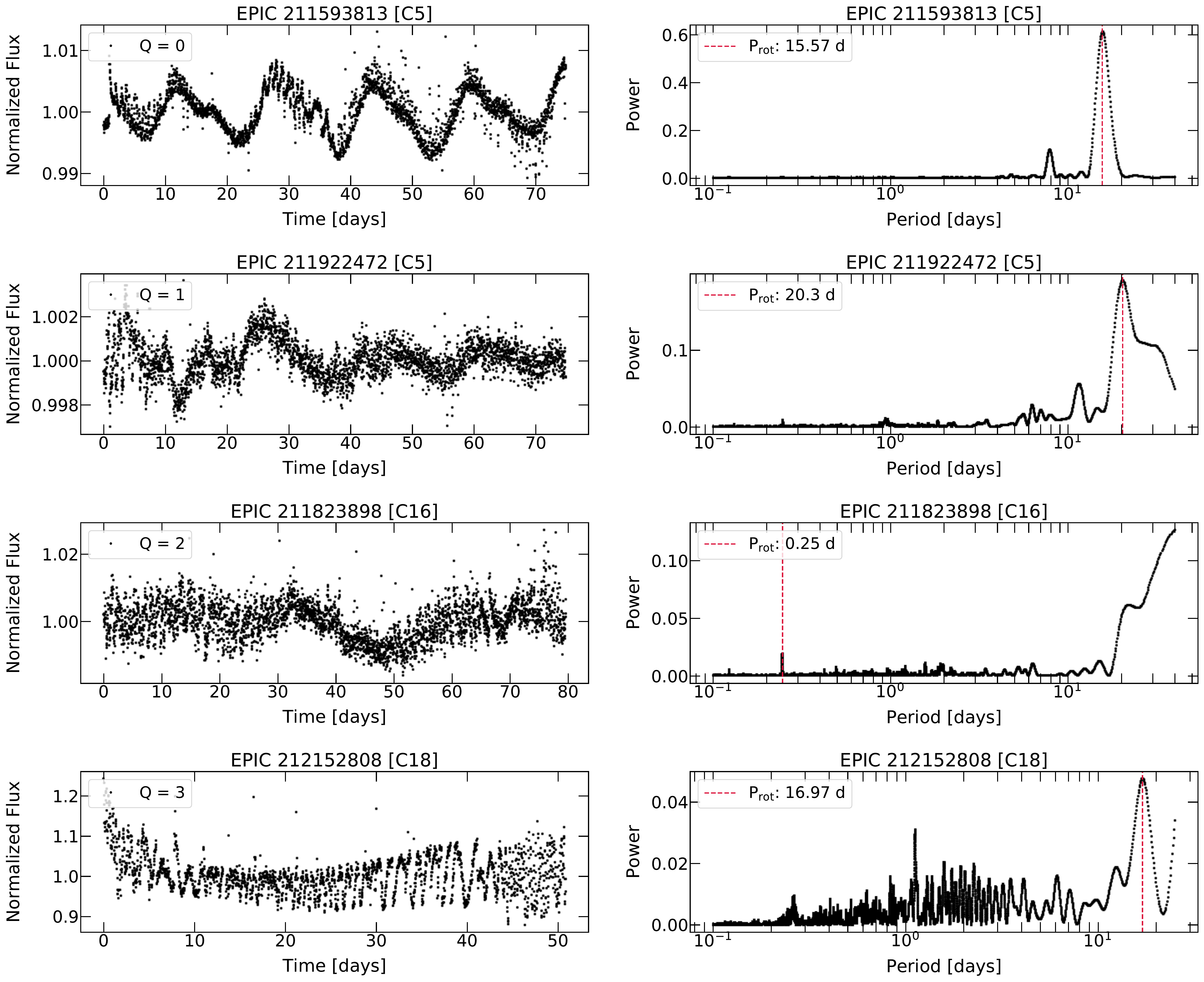}
    \caption{Examples of quality flag  designations. Light curves are on the left, with the corresponding periodogram on the right. The position of the detected period is indicated by the red dashed line. \textit{Top row:} Q = 0. There is a clear periodic modulation in the light curve and a strong, sharp peak in the periodogram. \textit{Second row:} Q = 1. There is evidence of modulation in the light curve, but the structure in the first half of the light curve makes it difficult to confidently claim that this \prot = 20 d.  \textit{Third row:} Q = 2. The light curve shows evidence for a long-term periodic trend likely due to systematics. This undermines our confidence in any \prot\ detected in the periodogram. \textit{Bottom row:} Q = 3. Systematics completely dominate the light curve structure. }
    \label{fig:qfex}
\end{figure*}



\end{document}